\documentclass[12pt,a4paper]{article}
\usepackage{jheppub}

\usepackage{etoolbox}
    \makeatletter
    \patchcmd{\maketitle}{\@fpheader}{}{}{}
    \makeatother

\usepackage[utf8]{inputenc}
\usepackage{lmodern}
\usepackage{exscale}

\usepackage{amsmath}
\usepackage{amssymb}
\usepackage{mathtools}

\usepackage{mathrsfs}

\usepackage{relsize,exscale}
\usepackage{tensor}

\usepackage[]{subdepth}

\newcommand{\scr}{\scriptscriptstyle}

\allowdisplaybreaks

\usepackage{graphicx}

\usepackage[dvipsnames]{xcolor}
\hypersetup{
    colorlinks,
    linkcolor={black},
    citecolor={black},
    urlcolor={black}
}

\title{Removing spurious degrees of freedom \\
from EFT of gravity}

\author[a]{Dra\v{z}en Glavan,}
\emailAdd{glavan@fzu.cz}
\author[b,c]{Shinji Mukohyama,}
\emailAdd{shinji.mukohyama@yukawa.kyoto-u.ac.jp}
\author[d]{Tom Zlosnik}
\emailAdd{thomas.zlosnik@ug.edu.pl}
\affiliation[a]{CEICO, FZU --- 
    Institute of Physics of the Czech Academy of Sciences,
	\\
	Na Slovance 1999/2, 182 21 Prague 8, Czech Republic}
\affiliation[b]{Center for Gravitational Physics and Quantum Information, Yukawa Institute for Theoretical Physics, Kyoto University, 606-8502, Kyoto, Japan}
\affiliation[c]{Kavli Institute for the Physics and Mathematics of the Universe (WPI), The University of Tokyo Institutes for Advanced Study (UTIAS), The University of Tokyo, Kashiwa, Chiba 277-8583, Japan}
\affiliation[d]{Institute of Theoretical Physics and Astrophysics,  University of Gda\'{n}sk, \\
ul.~Wita Stwosza 57, 80-308 Gda\'{n}sk, Poland}
\abstract{
In the effective field theory approach to gravity, the Lagrangian density for 
general relativity is supplemented by generally covariant terms of higher order 
in the Riemann tensor and its derivatives. At face value, these terms will result 
in higher derivative equations of motion and additional degrees of freedom 
beyond those of general relativity. This is seemingly at odds with the goal of 
effective field theory which is to encode ultraviolet corrections
in terms of existing low energy degrees of freedom. Here we apply an 
action-based procedure for the removal of spurious degrees of freedom for the 
case of general relativity supplemented by a term cubic in the Riemann 
tensor. 
To the order we work in, the resulting reduced theory takes the form of 
minimally modified gravity, which is a class of modified
gravity theories that propagate just the massless spin-$2$ graviton, but exhibit a preferred 
frame at short distances due to a deformation of time diffeomorphisms.
Our work suggests that this class of theories can be understood as an 
effective
theory of gravity, capturing effects of unknown ultraviolet physics.
}

\begin{document}
\begin{flushright} {\footnotesize YITP-24-117, IPMU24-0037}  \end{flushright}

\notoc
\maketitle
\titlepage

%\vskip+2cm

%%%%%%%%%%%%%%%%%%%%%%%%%%%%%%%%%%%%%%%%%%
%%%%%%%%%%%%%%%%%%%%%%%%%%%%%%%%%%%%%%%%%%
%%%%%%%%%%%%%%%%%%%%%%%%%%%%%%%%%%%%%%%%%%
%%%	INTRODUCTION
%%%%%%%%%%%%%%%%%%%%%%%%%%%%%%%%%%%%%%%%%%
%%%%%%%%%%%%%%%%%%%%%%%%%%%%%%%%%%%%%%%%%%
%%%%%%%%%%%%%%%%%%%%%%%%%%%%%%%%%%%%%%%%%%
\section{Introduction}
\label{sec: Introduction}

Pure general relativity is finite at one loop as a quantum theory in four spacetime dimensions, 
as proven by `t Hooft and Veltman~\cite{tHooft:1974toh}. This is essentially a 
consequence of the fact that both independent curvature-squared counterterms can be 
absorbed at one loop order by perturbative redefinition of the metric field
(perturbative disformal transformation). 
However, at two loops this is no longer true, as was shown by Goroff and 
Sagnotti~\cite{Goroff:1985sz,Goroff:1985th},
and later confirmed by van de Ven~\cite{vandeVen:1991gw}.
Some of the two-loop divergences take the form of Riemann tensor-cubed, which 
cannot be absorbed by analogous metric field redefinition. 
Thus, the higher derivative counterterm is necessary, and the theory is predictable
only in the effective theory sense, and has to be treated as such. 
At higher loop orders new higher derivative
counterterms need to be included.

The presence of higher derivative terms in the effective theory
of gravity will result in higher derivative equations of motion.
This is seemingly at odds with the purpose of the effective theory, 
which is supposed to encode ultraviolet corrections 
to the low energy propagating degrees of freedom, which is
just the massless spin-$2$ graviton with two physical polarizations. 
Lovelock's theorem~\cite{Lovelock:1971yv} states that the most
general diffeomorphism-invariant second-order equation of motion of
the form~$E_{\mu\nu} \!=\! 0$~for a 
four-dimensional theory of gravity described in terms of the metric only
is the Einstein equation with a cosmological 
constant, 
that descends from the Hilbert action,
and that propagates just the massless spin-2 graviton.
Diffeomorphism-invariant metric theories
with higher curvature actions tend to propagate additional degrees 
of freedom, apart from the massless spin-$2$ graviton. 
Moreover, degrees of freedom associated to higher derivatives will typically be afflicted 
by Ostrogradsky instabilities~\cite{Woodard:2015zca} that generically destabilize the theory. 
An exception to this is~$f(R)$ theory, that does not suffer from Ostrogradsky
instabilities, but propagate a scalar in addition to the massless spin-$2$ graviton. The curvature-squared
theory of gravity due to Stelle~\cite{Stelle:1976gc,Stelle:1977ry} is perturbatively 
renormalizable as a quantum theory, but propagates additional unwanted degrees of 
freedom. This is also the case with the local effective theory of gravity truncated at any 
particular order, that will seemingly
propagate more and more degrees of freedom the higher the order of truncation,
instead of just the massless spin-$2$ graviton as it is meant to.

Here we take the point of view that higher derivative terms in the effective 
theory of gravity result from some procedure of  integrating out the 
ultraviolet sector of some generally covariant theory. Integrating out degrees of
freedom generally results in an effective theory described by a non-local effective action. 
By construction, this effective theory propagates less degrees of freedom than the 
starting theory. However, nonlocal theories are difficult to work with, and a local
derivative expansion is often employed, assuming it is applied only below the
scale characteristic for ultraviolet physics. This derivative expansion results in an infinite series of 
local higher derivative terms making up the effective action, which is still equivalent
to its nonlocal counterpart. However, truncating this expansion at a finite order introduces 
artifacts of spurious degrees of freedom associated with higher derivatives. This 
precludes the truncated local effective theory to be taken at face 
value~\cite{Burgess:2014lwa}.
It should be noted that these spurious degrees of freedom are not related 
to the degrees of freedom that are integrated out~\cite{Simon:1990ic}.

Extracting reliable predictions from higher derivative local effective theories first
requires us to deal with the issue of spurious degrees of freedom. It is commonplace
to restrict ourselves to looking for strictly perturbative solutions of the theory, that take
the form of a power series where the leading term is a solution of the 
uncorrected theory. The corrections are then evaluated as perturbatively, which is
a process that cannot introduce additional degrees of freedom.
This approach has been applied to gravity to 
infer the perturbative stability of flat space~\cite{Simon:1990jn},
and to work out corrections
to black hole solutions and cosmology,
e.g.~\cite{Goon:2016mil,Calmet:2024pve}. 
While this method
is legitimate, it is somewhat restrictive, as it does not allow for resummations, i.e.~for
solving the effective theory exactly which might be called for in certain regimes
(e.g.~small effects over very long times can drive the evolution far away
from the uncorrected one~\cite{Parker:1993dk,Frob:2013ht}).
It is also somewhat unsatisfactory from a conceptual point of view, as we might gain 
better insight into the full theory by having an effective theory manifestly expressed
in terms of low energy degrees of freedom. In order to define a healthy theory it is 
useful to excise spurious degrees of freedom explicitly from the effective theory, i.e.~to 
reduce the derivative order of the theory.

Perturbative reduction of derivative order can be performed at the level of equations
of motion by applying lower order equations of motion to higher time derivatives 
terms in order to trade them for spatial derivatives or potential 
terms~\cite{Simon:1990jn,Parker:1993dk,Frob:2013ht}. 
Recently several closely related iterative numerical methods have been 
discussed~\cite{Gao:2023obj}
as order reduction methods for dealing with higher derivatives.
These procedure definitively
preserve strictly perturbative solutions of the theory. However, they can 
also exhibit
ambiguities: there are typically free parameters to choose in the reduced equations
such that they all reproduce one and the same solution to a given order of the truncation,
but differ when solved exactly. In general only specific choices of these parameters lead 
to equations of motion that descend from the action principle, which is highly preferable
as this guarantees that exact properties, such as e.g.~energy conservation, are preserved
in the truncated theory. Thus, it is preferable to perform the reduction of derivative order
at the level of the action. An action based algorithm for 
accomplishing this grounded in the canonical (Hamiltonian) formalism was
laid out in~\cite{Glavan:2017srd}, based on earlier seminal
works by Jaen, Llosa, and Molina~\cite{Jaen:1986iz}, and by Eliezer
and Woodard~\cite{Eliezer:1989cr}, 
that introduced the concept of perturbative constraints for reducing
derivative order and removing spurious degrees of freedom.
Utilizing the canonical formalism makes it possible to identify the effective second-class
constraint relations that can unambiguously be solved for to eliminate the 
spurious degrees of freedom. These effective constraints result from the requirement
that the solution space of the higher derivative theory is restricted to a subset
of solutions that are formally depend analytically on the coupling constants of 
higher derivative terms. This removes solutions associated to spurious degrees of freedom.

It was shown in~\cite{Glavan:2017srd} that the reduction algorithm works for theories 
without local symmetries, i.e.~theories without first-class constraints in the Dirac
classification~\cite{DiracBook}. However, it has remained an open question how well the
algorithm performs for theories with local symmetries such as spacetime diffeomorphism symmetry. Here we check whether the algorithm is applicable to gravity, 
at least at the first non-trivial order. We will consider the simplest non-trivial example of cubic gravity,
\begin{equation}
S[\boldsymbol{g}_{\mu\nu}] = \int\! d^{4\!}x \, \sqrt{ - \boldsymbol{g} } \, \biggl[
	\frac{\boldsymbol{R} - 2 \Lambda}{ \kappa^2}
	+ \alpha \kappa^2 {\boldsymbol{R}^{\mu\nu}}_{\rho\sigma} 
		{\boldsymbol{R}^{\rho\sigma}}_{\alpha\beta} {\boldsymbol{R}^{\alpha\beta}}_{\mu\nu}
	\biggr]
	\, .
\label{action}
\end{equation}
Here $\boldsymbol{g}_{\mu\nu}$ stands for the spacetime metric
with the mostly positive signature, with
$\boldsymbol{g} \!=\! {\rm det}(\boldsymbol{g}_{\mu\nu})$ its determinant,
$\Lambda$ is the cosmological constant, and~$\kappa^2\!=\! 16 \pi G_{\scr N}$ is the rescaled Newton's constant. For the Riemann tensor ${\boldsymbol{R}^{\mu\nu}}_{\rho\sigma}$ we adopt the conventions of \cite{Wald:1984rg}. The coupling constant~$\alpha$ is taken to be a formally small parameter,~$\alpha\!\ll\!1$.
Thus the cubic part of the action presents a correction to general relativity 
that cannot be removed by local field redefinitions and thus pushed 
to higher orders. 
There might be a way to absorb the Riemann-cubed term into the metric using
nonlocal field redefinition~\cite{Krasnov:2009ip,Krasnov:2009ik}, however, 
such nonlocal transformations should be approached with some care~\cite{Glavan:2024fsu}.

We should note that our treatment of the theory in~(\ref{action}) is different
from the recently proposed Einsteinian Cubic Gravity~\cite{Bueno:2016xff}
and its generalizations~\cite{Hennigar:2017ego,Arciniega:2018fxj}, that
would consider all terms in the action on equal footing (in addition to a number
of other cubic curvature invariants). Such theories seem
to propagate only the massless spin-$2$ graviton at the level of linearized perturbations around 
maximally symmetric backgrounds. However, despite this it does propagate additional 
degrees of freedom,
which is revealed even in the linear spectrum when considered around more general
backgrounds~\cite{Bueno:2023jtc,BeltranJimenez:2020lee,DeFelice:2023vmj,BeltranJimenez:2023mxp}. This restricts
Einsteinian Cubic Gravity to an effective theory valid on scales over which the 
additional degrees of freedom cannot be excited.
Our treatment of the Riemann-cubed term in~(\ref{action}) is rather as 
an effective local correction to the Einstein-Hilbert term, and we will seek its 
description with the spurious degrees of freedom removed, and with equations of
motion of second order in time derivatives.

Our reduction procedure will, by construction, select a preferred frame at short distances, as it singles 
out higher time
derivatives that are related to the number of independent initial conditions, i.e.~to 
the number of propagating degrees of freedom. 
Having that in mind, the resulting effective theory of gravity must preserve
spatial diffeomorphism invariance, and also must preserve the number of degrees of
freedom of general relativity. Therefore the resulting theory must 
 necessarily have the structure 
of {\it minimally modified gravity}~\cite{Lin:2017oow,Mukohyama:2019unx}
--- a modification of general relativity that (i) maintains spatial diffeomorphisms,
(ii) propagates two degrees of freedom, and (iii) maintains the on-shell 
first-class constraint structure of general relativity. Our computations indeed 
show this up to the order of expansion in $\alpha$. It should be noted that Ho\v{r}ava-Lifshitz gravity~\cite{Horava:2009uw}
shares property (i), but propagates three degrees of freedom so that it does not 
exhibit (ii) and (iii).

The structure of the paper is as follows: in Section \ref{sec: ADM decomposition and the
extended action} we present the Arnowitt-Deser-Misner decomposition of fields and the 
construction of the extended action. In 
Sec.~\ref{sec: Perturbative reduction of derivative order} we perform the perturbative 
reduction of derivative order of the action and cast it in canonical form. 
In Section \ref{sec: Simplifications and analysis of reduced canonical action} we simplify the 
canonical action by means of field redefinitions. Additionally, the algebra
of first-class constraints is found, and it is demonstrated that the resulting theory cannot be 
reduced to general relativity by way of canonical transformations. In 
Section~\ref{sec: Discussion} we discuss our results and present our conclusions. 
Additionally, Appendix \ref{sec: Additional mathematical details} contains a number of useful 
mathematical identities useful to the calculations in the main body in the paper, whilst 
Appendix \ref{sec:gravitational_waves} derives the propagation of gravitational waves on 
maximally-symmetric backgrounds for the order reduced theory.

%%%%%%%%%%%%%%%%%%%%%%%%%%%%%%%%%%%%%%%%%%
%%%%%%%%%%%%%%%%%%%%%%%%%%%%%%%%%%%%%%%%%%
%%%%%%%%%%%%%%%%%%%%%%%%%%%%%%%%%%%%%%%%%%
%%%	ADM DECOMPOSITION AND THE EXTENDED ACTION
%%%%%%%%%%%%%%%%%%%%%%%%%%%%%%%%%%%%%%%%%%
%%%%%%%%%%%%%%%%%%%%%%%%%%%%%%%%%%%%%%%%%%
%%%%%%%%%%%%%%%%%%%%%%%%%%%%%%%%%%%%%%%%%%
\section{ADM decomposition and the extended action}
\label{sec: ADM decomposition and the extended action}

The number of local propagating degrees of freedom of a theory is in correspondence
with the number of initial data that must be specified on some initial Cauchy surface
when formulating an initial value problem. The formalism particularly adapted to 
identifying these degrees of freedom is Dirac's canonical (Hamiltonian) formalism for
theories with constraints~\cite{DiracBook},
as it makes counting the number of degrees of freedom manifest. This is also
the formalism of choice for our procedure of action-based derivative reduction, 
i.e.~for removing spurious degrees of freedom associated to higher derivatives.
In this section we derive the extended action for the theory in~(\ref{action}),
which is the point of departure between the standard canonical formulation and the
reduction procedure we consider in the following section.

The first step towards the canonical formulation of the gravitational action
in~(\ref{action}) is to choose a particular time slicing, and to 
decompose indices into temporal and spatial ones accordingly (the
latter ones denoted by lowercase Latin letters). This is accomplished 
by the Arnowitt-Deser-Misner
(ADM) decomposition~\cite{Arnowitt:1962hi} that writes some of the (inverse) metric 
components,
\begin{equation}
\boldsymbol{g}^{00} = - \frac{1}{N^2} \, ,
\qquad \qquad
\boldsymbol{g}_{0i} = N_i \, ,
\qquad \qquad
\boldsymbol{g}_{ij} = g_{ij} \, ,
\label{ADMmetric1}
\end{equation}
in terms of ADM variables: lapse~$N$, shift~$N_i$, and the spatial metric~$g_{ij}$.
This induces the decomposition of the remaining (inverse) metric components,
\begin{equation}
\boldsymbol{g}_{00} = - N^2 + N_i N^i \, ,
\qquad \qquad
\boldsymbol{g}^{0i} = \frac{N^i}{N^2} \, ,
\qquad \qquad
\boldsymbol{g}^{ij} = g^{ij} - \frac{N^i N^j}{N^2} \, .
\label{ADMmetric2}
\end{equation}
where indices on the ADM variables are raised by the spatial metric,
$g_{ij} g^{jk} = \delta_i^k$, $N^i = g^{ij} N_j$. The metric determinant
is decomposed accordingly as~$\sqrt{-\boldsymbol{g}} = N \sqrt{g}$.
Note that the lapse is not allowed to vanish,~$N \!\neq\! 0$.
The ADM decomposition of Christoffel symbols is given by,
\begin{align}
\boldsymbol{\Gamma}^0_{00} ={}&
	\frac{1}{N} \Bigl( \dot{N} - N^i N^j K_{ij} + N^i \nabla_i N \Bigr) \, ,
\\
\boldsymbol{\Gamma}^0_{0i} ={}&
	- \frac{1}{N} \Bigl( N^j K_{ij} - \nabla_i N \Bigr) \, ,
\\
\boldsymbol{\Gamma}^0_{ij} ={}&
	- \frac{1}{N} K_{ij} \, ,
\\
\boldsymbol{\Gamma}^i_{00} ={}&
	- \frac{N^i}{N} \dot{N}
	+ g^{ij} \dot{N}_j
	+ \frac{N^i N^j}{N} \Bigl( N^k K_{jk} - \nabla_j N \Bigr)
	+ N \nabla^i N
	- N^j \nabla^i N_j \, ,
\\
\boldsymbol{\Gamma}^i_{0j} ={}&
	\frac{N^i}{N} \Bigl( N^k K_{jk} - \nabla_j N \Bigr)
	- N {K^i}_j + \nabla_j N^i \, ,
\\
\boldsymbol{\Gamma}^i_{jk} ={}&
	\Gamma^i_{jk} + \frac{N^i}{N} K_{jk} \, .
\label{ExtrinsicCurvature}
\end{align}
where~$\nabla_i$ denotes the three-dimensional covariant derivative
with respect to the spatial metric~$g_{ij}$ and the corresponding Christoffel 
symbols on spatial 
slices,~$\Gamma^{k}_{ij}\!=\! \frac{1}{2} g^{kl} \bigl( \partial_i g_{jl} \!+\! \partial_j g_{il} \!-\! \partial_l g_{ij} \bigr)$. Checking the decompositions above, and the rest of
the decompositions in this section is greatly streamlined by the use of 
{\it Cadabra}~\cite{Peeters:2007wn,Peeters:2006kp,Peeters:2018dyg},
a symbolic computer algebra system for tensor manipulation.
The extrinsic curvature tensor is defined as,
\begin{equation}
K_{ij} = - \frac{1}{2N} \Bigl( \dot{g}_{ij} - \nabla_i N_j - \nabla_j N_i \Bigr) \, .
\end{equation}
The decompositions and definitions in~(\ref{ADMmetric1})--(\ref{ExtrinsicCurvature})
suffice to decompose the Riemann tensor, most conveniently given in terms
of four independent  components,
\begin{subequations}
\begin{align}
{\boldsymbol{R}^{0i}}_{0j} ={}&
	{F^i}_j 
	- \frac{1}{N} {L^i}_{jk} N^k \, ,
\label{ADMriemann1}
\\
{\boldsymbol{R}^{0i}}_{jk} ={}&
	\frac{1}{N} {L^i}_{jk} \, ,
\label{ADMriemann2}
\\
{\boldsymbol{R}^{ij}}_{0k} ={}&
	- 2 N^{[i} {F^{j]}}_k
	+ \frac{2}{N} N^{[i} {L^{j]}}_{kl} N^l
	- N {L_k}^{ij}
	- {Q^{ij}}_{kl} N^l \, ,
	\qquad
\label{ADMriemann3}
\\
{\boldsymbol{R}^{ij}}_{kl} ={}&
	{Q^{ij}}_{kl}
	- \frac{2}{N} N^{[i} {L^{j]}}_{kl} \, ,
\label{ADMriemann4}
\end{align}
\label{ADMriemann}%
\end{subequations}
Here we have introduced definitions of three ADM tensors,
\begin{subequations}
\begin{align}
&
F_{i j} = - \frac{1}{N} \dot{K}_{ij}
	- K_{ik} {K^k}_j
	+ \frac{N^k}{N} \nabla_k K_{ij}
	+ \frac{2}{N} K_{k(i} \nabla_{j)} N^k
	- \frac{1}{N} \nabla_i \nabla_j N \, ,
\\
& {Q^{ij}}_{kl} = 2 {K^i}_{[k} {K_{l]}}^j
	+ {R^{ij}}_{kl} \, ,
\qquad \qquad
{L^i}_{jk} = 2 \nabla_{[k} {K_{j]}}^i \, ,
\end{align}
\end{subequations}
and shorthand notation for their contractions,
\begin{equation}
F = {F^i}_i \, ,
\qquad \quad
{Q^i}_j = {Q^{ik}}_{jk} \, ,
\qquad \quad
Q = {Q^i}_i \, ,
\qquad \quad
L_i = {L^j}_{ji} \, .
\end{equation}
The tensor~$F_{ij}$, introduced in~\cite{Buchbinder:1987vp}, can be thought of as 
capturing the second time derivative of the 
metric,~${Q^{ij}}_{kl}$ as capturing the square of the first time derivative of the metric, 
and~${L^{i}}_{jk}$ the gradient of the first time derivative.
The ADM decomposition of the cubic term in the action~(\ref{action})
now follows,
\begin{equation}
{\boldsymbol{R}^{\mu\nu}}_{\rho\sigma}  {\boldsymbol{R}^{\rho\sigma}}_{\alpha\beta} {\boldsymbol{R}^{\alpha\beta}}_{\mu\nu}
=
	4 {F^i}_j \Bigl( 2 {F^j}_k {F^k}_i
		- 3 {L_i}^{kl} {L^j}_{kl} \Bigr)
	+ {Q^{ij}}_{kl} \Bigl( {Q^{kl}}_{mn} {Q^{mn}}_{ij} 
		- 6 {L_m}^{kl} {L^m}_{ij} \Bigr)
	\, .
\end{equation}
The decomposition of the Ricci tensor is inferred by contracting the appropriate
components in~(\ref{ADMriemann}),
\begin{subequations}
\begin{align}
{\boldsymbol{R}^0}_0 ={}&
	F - \frac{1}{N} N^i L_i \, ,
\\
{\boldsymbol{R}^0}_i ={}&
	- \frac{1}{N} L_i \, ,
\\
{\boldsymbol{R}^i}_j ={}&
	{F^i}_j + {Q^i}_j + \frac{1}{N} N^i L_j \, ,
\\
{\boldsymbol{R}^i}_0 ={}&
	- \Bigl( F - \frac{1}{N} N^j L_j \Bigr) N^i
	+ \Bigl( {F^i}_j + {Q^i}_j \Bigr) N^j
	+ N L^i \, ,
\end{align}
\end{subequations}
while contracting these gives the Ricci scalar,
\begin{equation}
\boldsymbol{R} = 2F + Q \, .
\end{equation}
However, the second derivative of the metric contained in~$F$ in the Ricci scalar 
decomposition above can be partially integrated to yield the ADM decomposed action,
\begin{align}
\MoveEqLeft[3]
S \bigl[ N, N_i, g_{ij} \bigr] 
	= 
	\int\! d^{4\!} x \, N \sqrt{g} \biggl[
		\frac{1}{\kappa^2} \Bigl( {K^i}_j {K^j}_i - K^2 + R - 2 \Lambda  \Bigr)
\\
&
	+ 4 \alpha \kappa^2 {F^i}_j \Bigl( 2 {F^j}_k {F^k}_i
		- 3 {L_i}^{kl} {L^j}_{kl} \Bigr)
	+ \alpha \kappa^2 {Q^{ij}}_{kl} \Bigl( {Q^{kl}}_{mn} {Q^{mn}}_{ij} 
		- 6 {L_m}^{kl} {L^m}_{ij} \Bigr)
	\biggr] \, .
\nonumber 
\end{align}

We proceed by constructing the extended action, where time derivatives are promoted 
to independent velocity fields,
\begin{equation}
K_{ij} \longrightarrow \mathcal{K}_{ij} \, ,
\qquad \qquad
F_{ij} \longrightarrow \mathcal{F}_{ij} \, ,
\end{equation}
and accordingly,
\begin{equation}
{L^i}_{jk} \longrightarrow {\mathcal{L}^i}_{jk} = 2 \nabla_{[k} { \mathcal{K}_{j]}}^i \, ,
\qquad \quad
{Q^{ij}}_{kl} \longrightarrow {\mathcal{Q}^{ij}}_{kl}
	= 2 {\mathcal{K}^i}_{[k} {\mathcal{K}_{l]}}^j
	+ {R^{ij}}_{kl} \, , 
\label{DependentTensors}
\end{equation}
and where accompanying Lagrange multipliers~$\pi_{ij}$ and~$\rho_{ij}$ are
are introduced to ensure on-shell equivalence,
\begin{align}
\MoveEqLeft[3]
\mathcal{S} \bigl[ N, N_i, g_{ij}, \mathcal{K}_{ij}, \mathcal{F}_{ij}, \pi^{ij}, \rho^{ij} \bigr] 
= \int\! d^{4\!} x \, \Biggl\{
	N \sqrt{g} \biggl[
		\frac{1}{\kappa^2} \Bigl( {\mathcal{K}^i}_j { \mathcal{K}^j}_i - \mathcal{K}^2 + R - 2 \Lambda  \Bigr)
\nonumber \\
&
	+ 4 \alpha \kappa^2 {\mathcal{F}^i}_j \Bigl( 2 {\mathcal{F}^j}_k {\mathcal{F}^k}_i
		- 3 {\mathcal{L}_i}^{kl} {\mathcal{L}^j}_{kl} \Bigr)
	+ \alpha \kappa^2 {\mathcal{Q}^{ij}}_{kl} 
        \Bigl( {\mathcal{Q}^{kl}}_{mn} {\mathcal{Q}^{mn}}_{ij} 
		- 6 {\mathcal{L}_m}^{kl} {\mathcal{L}^m}_{ij} \Bigr)
	\biggr] 
\nonumber \\
&
	+ \pi^{ij} \Bigl( \dot{g}_{ij} - 2 \nabla_{(i} N_{j)} + 2 N \mathcal{K}_{ij} \Bigr)
	+ \rho^{ij} \Bigl( \dot{\mathcal{K}}_{ij}
		+ N \mathcal{K}_{ik} {\mathcal{K}^k}_j
		- N^k \nabla_k \mathcal{K}_{ij}
\nonumber \\
&
	- 2 \mathcal{K}_{k(i} \nabla_{j)} N^k
	+ \nabla_i \nabla_j N + N \mathcal{F}_{ij} \Bigr)
\Biggr\} 
\, .
\label{ExtendedAction}
\end{align}
The action above is now in first order form, but it is not the canonical action yet.
The canonical action would be obtained by following these steps: (i) obtain the
equation resulting from varying with respect to~$\mathcal{F}_{ij}$, (ii) solve for
as many components of~$\mathcal{F}_{ij}$ as possible, and (iii) plug those on-shell
solutions into the extended action above. Such obtained canonical action would
treat both the Einstein-Hilbert and the Riemann-cubed parts on equal footing, and
would be the canonical formulation of the genuinely higher derivative system.
Henceforth we do not follow this route. Rather, we 
make an additional assumption about the nature of the Riemann-cubed term ---
that it has arisen by a local derivative expansion and the truncation of the
full effective action. This assumption prevents the Riemann-cubed part of
the action from generating additional degrees of freedom that would arise when it were treated on the 
same footing as the Einstein-Hilbert part. In the following section we discuss
how to implement this requirement mathematically, and how it modifies the
procedure of obtaining the canonical action from the extended action.

%%%%%%%%%%%%%%%%%%%%%%%%%%%%%%%%%%%%%%%%%%
%%%%%%%%%%%%%%%%%%%%%%%%%%%%%%%%%%%%%%%%%%
%%%%%%%%%%%%%%%%%%%%%%%%%%%%%%%%%%%%%%%%%%
%%%	PERTURBATIVE REDUCTION OF DERIVATIVE ORDER
%%%%%%%%%%%%%%%%%%%%%%%%%%%%%%%%%%%%%%%%%%
%%%%%%%%%%%%%%%%%%%%%%%%%%%%%%%%%%%%%%%%%%
%%%%%%%%%%%%%%%%%%%%%%%%%%%%%%%%%%%%%%%%%%
\section{Perturbative reduction of derivative order}
\label{sec: Perturbative reduction of derivative order}

We make one fundamental assumption about our system given by the action~(\ref{action}),
and consequently by the extended action~(\ref{ExtendedAction}): that all the fields
have to formally be analytic functions of the coupling constant~$\alpha$ on-shell. This implies that
all the quantities can formally be expanded as a power series 
in~$\alpha$, in accordance with the assumption of how the higher derivative 
terms arise in the first place --- as perturbative corrections. 
This assumption guarantees
that the reduced theory smoothly connects to the leading order
Einstein-Hilbert theory. In practice this means that we are not allowed
to divide by~$\alpha$, which precludes going from the extended 
action~(\ref{ExtendedAction}) directly to the canonical action.
Rather, the requirement of analyticity introduces effective second-class
constraints that allow us to solve and eliminate dynamical variables 
associated to spurious degrees of freedom.
We follow the algorithm laid out in~\cite{Glavan:2017srd}, that is based on earlier 
works~\cite{Jaen:1986iz,Eliezer:1989cr}. We work to the first relevant order in~$\alpha$,
meaning that we neglect terms of order~$\mathcal{O}(\alpha^2)$. This is implicit
in every equation of this section, and we omit denoting this explicitly. Our primary task
is to construct the canonical action where all the canonical variables are analytic functions
of~$\alpha$.

There are four equations descending from variation with respect to
canonical variables,
\begin{align}
\frac{\delta \mathcal{S}}{\delta \mathcal{\rho}^{ij}}
	={}&
	\dot{\mathcal{K}}_{ij}
		+ N \mathcal{K}_{ik} {\mathcal{K}^k}_j
		- N^k \nabla_k \mathcal{K}_{ij}
		- 2 \mathcal{K}_{k(i} \nabla_{j)} N^k
		+ \nabla_i \nabla_j N + N \mathcal{F}_{ij}
		\approx 0 
		\, ,
\label{EOM1}
\\
\frac{\delta \mathcal{S}}{\delta \mathcal{K}_{ij}}
    ={}&
    \! -
    \dot{\rho}^{ij}
    \!+
    \sqrt{g}
    \biggl[
    2 \bigl( N {\mathcal{K}_k}^{(i}
    \!-\!
    \nabla_k N^{(i}
    \bigr)
    \frac{ \rho^{j)k} }{ \sqrt{g} } 
    +
    \nabla_k \Bigl( N^k \frac{\rho^{ij} }{ \sqrt{g} } \Bigr)
    +
    2N
    \Bigl(
    \frac{ \pi^{ij} }{ \sqrt{g} }
    \!+\!
    \frac{ \mathcal{K}^{ij} \!-\! g^{ij} \mathcal{K} }{\kappa^2} 
    \Bigr)
    \biggr]
\nonumber \\
& - 12 \alpha \kappa^2 N \sqrt{g} \, 
        \biggl[
            4\nabla_l  {\mathcal{L}_k}^{l(i} \mathcal{F}^{j)k} 
            +
            \mathcal{Q}^{klm(i} {\mathcal{Q}^{j)n}}_{kl} \mathcal{K}_{mn}
\nonumber \\
&   \hspace{3cm}
    -
    2 \mathcal{L}^{mk(i} {\mathcal{L}_m}^{j)l} \mathcal{K}_{kl}
    +
    \frac{2}{N} 
        \nabla_m \bigl( N {\mathcal{Q}_{kl}}^{m(i} \mathcal{L}^{j)kl} \bigr)
    \biggr]
    \approx 0
    \, ,
\label{EOM2}
\\
\frac{\delta \mathcal{S}}{\delta \mathcal{\pi}^{ij}}
	={}&
	\dot{g}_{ij} - 2 \nabla_{(i} N_{j)} + 2 N \mathcal{K}_{ij}
	\approx 0
	\, ,
\label{EOM3}
\\
\frac{\delta \mathcal{S}}{\delta g_{ij}}
	={}&
    -
    \dot{\pi}^{ij}
    +
    \frac{N \sqrt{g} }{\kappa^2}  \,
    \biggl[
        -
        2 \mathcal{K}^{k(i} {\mathcal{K}^{j)}}_{k} 
        +
        2 \mathcal{K}^{ij} \mathcal{K}
        -
        G^{ij}
        +
        \frac{ g^{ij} }{2} 
            \bigl( {\mathcal{K}^i}_j { \mathcal{K}^j}_i - \mathcal{K}^2 - 2\Lambda \bigr)
        \biggr]
\nonumber \\
&
\hspace{-1.2cm}
    +
    \frac{ \sqrt{g}  }{\kappa^2}
		\Bigl( 
            \nabla^i \nabla^j N
            \!-\!
            g^{ij} \nabla^k \nabla_k N
            \Bigr)
    \!+\!
    \sqrt{g} \,
        \nabla_k \biggl[ 
        N^k \frac{ \pi^{ij} }{ \sqrt{g} }
        \!-\!
        2 \frac{ \pi^{k(i} }{ \sqrt{g} } N^{j)}
        \!+\!
	    \frac{\rho^{k(i}}{\sqrt{g}} \nabla^{j)} N
        \!-\!
        \frac{1}{2}
	    \frac{ \rho^{ij} }{ \sqrt{g} } \nabla^k N
\nonumber \\
&
\hspace{-1.2cm}
        \!-\!
    2
    \frac{ \rho^{kl} }{ \sqrt{g} } {\mathcal{K}^{(i}}_l N^{j)}
    \biggr]
    +
    \rho^{kl} \Bigl(
    2 {\mathcal{K}^{(i}}_{k} \nabla_{l} N^{j)}
    +
	N^{(i} \nabla^{j)} \mathcal{K}_{kl}
    - 
    N {\mathcal{K}^{(i}}_k {\mathcal{K}^{j)}}_l
    \Bigr)
    +
    \alpha \kappa^2 \sqrt{g} \,
    \mathscr{G}^{ij}  
    \approx 0
    \, ,
\label{EOM4}
\end{align}
where the lengthy tensor~$\mathscr{G}^{ij}$ is given in~(\ref{GijLengthy})
in the Appendix.
Then there are three equations descending from variations with respect to 
Lagrange multipliers/auxiliary variables that that generate algebraic equations
only,
\begin{align}
&
\frac{\delta \mathcal{S}}{ \delta N}
	=
	\sqrt{g} \biggl[
	2 \mathcal{K}_{ij} \frac{ \pi^{ij} }{ \sqrt{g} }
	+
	\bigl( \mathcal{K}_{ik} {\mathcal{K}^k}_j \!+\! \mathcal{F}_{ij}  \bigr) \frac{ \rho^{ij} }{ \sqrt{g} } 
	+ \!
	\nabla_i \nabla_j \Bigl( \frac{ \rho^{ij} }{ \sqrt{g} } \Bigr)
	\!+\!
	\frac{1}{\kappa^2} \Bigl( {\mathcal{K}^i}_j { \mathcal{K}^j}_i \!-\! \mathcal{K}^2 \!+\! R \!-\! 2 \Lambda  \Bigr)
\nonumber \\
&	\hspace{0.2cm}
	+ 4 \alpha \kappa^2 
    {\mathcal{F}^i}_j \Bigl( 2 {\mathcal{F}^j}_k {\mathcal{F}^k}_i
		- 3 {\mathcal{L}_i}^{kl} {\mathcal{L}^j}_{kl} \Bigr)
	+ \alpha \kappa^2 
    {\mathcal{Q}^{ij}}_{kl} \Bigl( {\mathcal{Q}^{kl}}_{mn} {\mathcal{Q}^{mn}}_{ij} 
		- 6 {\mathcal{L}_m}^{kl} {\mathcal{L}^m}_{ij} \Bigr)
	\biggr]
	\approx 0
	\, ,
\label{EOM5}
\\
&
\frac{\delta \mathcal{S}}{ \delta N_i }
	=
	\sqrt{g}
	\biggl[
	2 \nabla_{j} \Bigl( \frac{ \pi^{ij} }{ \sqrt{g} } \Bigr)
	+ 2 \nabla_{j} \Bigl( {\mathcal{K}^i}_{k} \frac{ \rho^{jk} }{ \sqrt{g} } \Bigr)
	- \frac{ \rho^{jk} }{ \sqrt{g} } \nabla^i \mathcal{K}_{jk}
	\biggr]
	\approx 0
	\, ,
\label{EOM6}
\\
&
\frac{ \delta \mathcal{S} }{ \delta \mathcal{F}_{ij} }
	=
	N \sqrt{g} \biggl[
	\frac{ \rho^{ij} }{ \sqrt{g} }
	+
	12 \alpha \kappa^2 \Bigl( 2 \mathcal{F}^{ik} {\mathcal{F}_k}^j
		- \mathcal{L}^{ikl} {\mathcal{L}^j}_{kl} \Bigr)
	\biggr]
	\approx 0
	\, .
\label{EOM7}
\end{align}
In the following we solve for algebraic equations maintaining perturbativity/analyticity
in~$\alpha$. This involves solving equations order by order starting from the leading order.
The algebraic equations change character when we approach them in this way.
If taken at face value, the variable~$\mathcal{F}_{ij}$, that appears cubically in the action, would be an auxiliary variable that we would in principle be able to solve for it. This is not the case when perturbativity in~$\alpha$ is invoked. That requirement
essentially turns~$\mathcal{F}_{ij}$ into a Lagrange multiplier because
it appears only linearly at leading order.

The analysis to follow benefits from defining shorthand 
notation for three tensors,%
\begin{subequations}
\begin{align}
&
\mathscr{K}_{ij} \equiv
    -
    \kappa^2 \biggl[
		\frac{ \pi_{ij} }{ \sqrt{g} }
		- \frac{g_{ij}}{2} \frac{ \pi }{ \sqrt{g} }
		\biggr]
  \, ,
\quad
{\mathscr{L}^i}_{jk} 
	\equiv  2 \nabla_{[k} {\mathscr{K}_{j]}}^i =
    \kappa^2 \biggl[
\delta^i_{[j} \nabla_{{k]}} 
		\Bigl(
		\frac{ \pi }{ \sqrt{g} }
		\Bigr)
- 
	2 \nabla_{[k} 
		\Bigl(
		\frac{ {\pi_{j]} }^i }{ \sqrt{g} }
		\Bigr)
    \biggr]
		\, , \label{calleq}
\\
&
{\mathscr{Q}^{ij}}_{kl} 
	\equiv
	2{\mathscr{K}^i}_{[k} {\mathscr{K}^j}_{l]}
	+
	{R^{ij}}_{kl} 
\nonumber \\
&   \hspace{1cm}
    = 2 \kappa^4 \biggl[
		\frac{ {\pi^i}_{[k} }{ \sqrt{g} }
		\frac{ {\pi^j}_{l]} }{ \sqrt{g} }
		- 
		\frac{ \delta^i_{[k} }{2} 
		\frac{ {\pi^j}_{l]} }{ \sqrt{g} }
		\frac{ \pi }{ \sqrt{g} }
		+
		\frac{ \delta^j_{[k} }{2}
		\frac{ {\pi^i}_{l]} }{ \sqrt{g} }
		\frac{ \pi }{ \sqrt{g} }
		+
		\frac{ \delta^i_{[k} \delta^j_{l]} }{4}
		\frac{ \pi }{ \sqrt{g} }
		\frac{ \pi }{ \sqrt{g} }
		\biggr]
	+
	{R^{ij}}_{kl}
	\, , \label{calqeq}
\end{align}
\label{KLQdefinitions}%
\end{subequations}
and their contractions,
\begin{subequations}
\begin{align}
\mathscr{K} \equiv{}& 
    {\mathscr{K}^i}_i
    =
	\frac{ \kappa^2 }{2} \frac{ \pi }{ \sqrt{g} }
    \, ,
\qquad
{\mathscr{Q}^i}_j
	\equiv
	{\mathscr{Q}^{ik}}_{jk}
	=
	- \kappa^4 \biggl[
		\frac{ {\pi^i}_{k} }{ \sqrt{g} }
		\frac{ {\pi^k}_{j} }{ \sqrt{g} }
		-
		\frac{1}{2}
		\frac{ {\pi^i}_{j} }{ \sqrt{g} }
		\frac{ \pi }{ \sqrt{g} }
		\biggr]
	+
	{R^i}_{j}
	\, ,
\\
{\mathscr{L}}_i
	\equiv{}&
	{\mathscr{L}^j}_{ji} 
	=
	\kappa^2 \nabla^{j} 
		\Bigl( \frac{ \pi_{ji} }{ \sqrt{g} } \Bigr)
		\, ,
\qquad
\mathscr{Q}
	\equiv
	{\mathscr{Q}^i}_i
	=
	- \kappa^4 \biggl[
		\frac{ {\pi^i}_{j} }{ \sqrt{g} }
		\frac{ {\pi^j}_{i} }{ \sqrt{g} }
		-
		\frac{1}{2}
		\frac{ \pi }{ \sqrt{g} }
		\frac{ \pi }{ \sqrt{g} }
		\biggr]
	+
	R
	\, .
\label{LiQdef}
\end{align}
\label{KLQcontractions}%
\end{subequations}
These will help keep the expressions considerably more compact.

%%%%%%%%%%%%%%%%%%%%%%%%%%%%%%%%%%%%%%%%%%
%%%%%%%%%%%%%%%%%%%%%%%%%%%%%%%%%%%%%%%%%%
%%%	EQUATIONS AT LEADING ORDER
%%%%%%%%%%%%%%%%%%%%%%%%%%%%%%%%%%%%%%%%%%
%%%%%%%%%%%%%%%%%%%%%%%%%%%%%%%%%%%%%%%%%%
\subsection{Equations at leading order}
\label{subsec: Equations at leading order}

This task begins with listing all the leading order equations descending from the 
extended action~(\ref{ExtendedAction}). These are captured by multiplying all the
equations by~$\alpha$, and writing all the expressions up to order~$\mathcal{O}(\alpha^2)$
implicitly. There are four equations descending from variation with respect to
canonical variables,
\begin{align}
&
\alpha
\frac{\delta \mathcal{S}}{\delta \mathcal{\rho}^{ij}}
	=
	\alpha \Bigl[ \dot{\mathcal{K}}_{ij}
		\!+ N \mathcal{K}_{ik} {\mathcal{K}^k}_j
		- N^k \nabla_k \mathcal{K}_{ij}
		- 2 \mathcal{K}_{k(i} \nabla_{j)} N^k
		\!+ \nabla_i \nabla_j N 
		+ N \mathcal{F}_{ij} \Bigr]
		\!\approx 0 \, ,
\label{AlphaEOM1}
\\
&
\alpha
\frac{\delta \mathcal{S}}{\delta \mathcal{K}_{ij}}
	=
	\alpha \sqrt{g} \biggl[
		- \frac{ \dot{\rho}^{ij} }{ \sqrt{g} }
		+ 2 \Bigl( N {{\mathcal{K}_k}^{(i} } - \nabla_k N^{(i}  \Bigr) \frac{\rho^{j)k} }{ \sqrt{g} } 
		+ \nabla_k \Bigl( N^k \frac{\rho^{ij} }{ \sqrt{g} } \Bigr)
\nonumber \\
&	\hspace{7cm}
	+ 2 N \frac{ \pi^{ij} }{ \sqrt{g} }
	+  \frac{ 2N }{\kappa^2} \bigl( \mathcal{K}^{ij} - g^{ij} \mathcal{K} \bigr)
	\biggr]
	\approx 0 \, ,
\label{AlphaEOM2}
\\
&
\alpha
\frac{\delta \mathcal{S}}{\delta \mathcal{\pi}^{ij}}
	=
	\alpha \Bigl[ \dot{g}_{ij} - 2 \nabla_{(i} N_{j)} + 2 N \mathcal{K}_{ij} \Bigr]
	\approx 0
	\, ,
\label{AlphaEOM3}
\\
&
\alpha
\frac{\delta \mathcal{S}}{\delta g_{ij}}
	=
    \alpha \biggl\{
    - \dot{\pi}^{ij}
    +
    \frac{N \sqrt{g}}{\kappa^2}
    \biggl[
    - 2 \mathcal{K}^{ik} {\mathcal{K}_k}^j
    + 2 \mathcal{K}^{ij} \mathcal{K}
    - R^{ij}
    + \frac{ g^{ij} }{ 2 }
        \Bigl(
        \mathcal{K}^{kl} \mathcal{K}_{kl} 
        \!-\! \mathcal{K}^2
        \!+\! R \!-\! 2\Lambda
        \Bigr)
    \biggr]
\nonumber \\
&   +
    \sqrt{g}
    \biggl[
    \frac{1}{\kappa^2}
    \Bigl(
    \nabla^i \nabla^j N
    -
    g^{ij} \nabla^k \nabla_k N
    \Bigr)
    - 
    2 \nabla_k \Bigl( \frac{ N^{(i}  \pi^{j)k}}{ \sqrt{g} } \Bigr)
    + 
    \nabla_k \Bigl( N^k \frac{\pi^{ij}}{ \sqrt{g} } \Bigr)
    \biggr]+ {\cal O}(\rho)
    \, .
	\biggr\} \, ,
\label{AlphaEOM4}
\end{align}
Then there are three equations descending from variations with respect to 
Lagrange multipliers/auxiliary variables that generate algebraic equations
only,
\begin{align}
\alpha
\frac{\delta \mathcal{S}}{ \delta N}
	={}&
	\alpha
	\sqrt{g} \biggl[
	2 \mathcal{K}_{ij} \frac{ \pi^{ij} }{ \sqrt{g} }
	+
	\bigl( \mathcal{K}_{ik} {\mathcal{K}^k}_j \!+\! \mathcal{F}_{ij}  \bigr) \frac{ \rho^{ij} }{ \sqrt{g} } 
	+ \!
	\nabla_i \nabla_j \Bigl( \frac{ \rho^{ij} }{ \sqrt{g} } \Bigr)
\nonumber \\
&	\hspace{2cm}
	+
	\frac{1}{\kappa^2} \Bigl( {\mathcal{K}^i}_j { \mathcal{K}^j}_i - \mathcal{K}^2 + R - 2 \Lambda  \Bigr)
	\biggr]
	\equiv
	- \alpha \mathcal{H}
	\approx 0
	\, ,
\label{AlphaEOM5}
\\
\alpha
\frac{\delta \mathcal{S}}{ \delta N_i }
	={}&
	\alpha
	\sqrt{g}
	\biggl[
	2 \nabla_{j} \Bigl( \frac{ \pi^{ij} }{ \sqrt{g} } \Bigr)
	+ 2 \nabla_{j} \Bigl( {\mathcal{K}^i}_{k} \frac{ \rho^{jk} }{ \sqrt{g} } \Bigr)
	- \frac{ \rho^{jk} }{ \sqrt{g} } \nabla^i \mathcal{K}_{jk}
	\biggr]
	\equiv
	- \alpha \mathcal{H}^i
	\approx 0
	\, ,
\label{AlphaEOM6}
\\
\alpha
\frac{ \delta \mathcal{S} }{ \delta \mathcal{F}_{ij} }
	={}&
	\alpha
	N \rho^{ij}
	\equiv
	- \alpha N \Phi^{ij}
	\approx 0
	\, .
\label{AlphaEOM7}
\end{align}
It is now clear that at leading order the latter three equations take
the form of constraints since one cannot solve for~$N$,~$N_i$,
nor~$\mathcal{F}_{ij}$. Thus~$\alpha\mathcal{H}$,~$\alpha \mathcal{H}^i$,
and~$\alpha \Phi^{ij}$, defined above, should be treated as constraints.
The first two appear as Hamiltonian and momentum constraints
as in any diffeomorphism invariant theory of gravity, while the additional constraint~(\ref{AlphaEOM7})
appears at leading order due to our requirement of analyticity in~$\alpha$.

The conservation of~$\alpha \mathcal{H}$ and~$\alpha \mathcal{H}^i$
does not generate any further constraints, but the conservation of~$\alpha\Phi^{ij}$
does,
\begin{equation}
\alpha \dot{\Phi}^{ij} \approx
	- 2 \alpha N \sqrt{g} \biggl[
	\frac{\pi^{ij}}{ \sqrt{g} }
	+ \frac{1}{\kappa^2} 
        \bigl( \mathcal{K}^{ij} \!-\! g^{ij} \mathcal{K} \bigr)
	\biggr]
	\equiv 
	\alpha N \Psi^{ij} 
	\approx 0
	\, .
\end{equation}
The conservation of the secondary constraint~$\Psi^{ij}$ then
generates another on-shell algebraic condition,
\begin{align}
\alpha \dot{\Psi}^{ij} \approx{}&
    \alpha \kappa^2 N \sqrt{g}
    \biggl[
    - 2 \frac{ \pi^{ik} }{ \sqrt{g} } 
        \frac{ {\pi_k}^j }{ \sqrt{g} } 
    + \frac{\pi}{ \sqrt{g} } \frac{ \pi^{ij} }{ \sqrt{g} } 
    + g^{ij} \Bigl(
        \frac{ \pi^{kl} }{ \sqrt{g} } \frac{ \pi_{kl} }{ \sqrt{g} } 
    - \frac{ 1 }{2}
        \frac{ \pi }{ \sqrt{g} } \frac{ \pi }{ \sqrt{g} } 
        \Bigr)
\nonumber \\
&    + \frac{2}{\kappa^4} \Bigl( G^{ij} + \mathcal{F}^{ij} 
        - g^{ij} \mathcal{F} + g^{ij} \Lambda \Bigr)
    \biggr]
    \approx 0 \, .
\end{align}
However, this condition does not amount to another constraint, but rather
it determines what the multiplier~$\mathcal{F}_{ij}$ is on-shell at leading order.
%In the language of Dirac's formalism for theories with 
%constraints,~$\Phi^{ij}$ and~$\Psi^{ij}$ are second-class constraints whose 
%nonvanishing Poisson bracket leads to the determination of the  Lagrange 
%multiplier~$\mathcal{F}_{ij}$ on-shell. This is the mathematical structure 
%stemming from the requirement of analyticity in~$\alpha$
%that allows to excise the degrees of freedom associated to higher derivatives.
Thus we can solve for~$\mathcal{K}_{ij}$ and~$\rho^{ij}$ from the two 
constraints,
\begin{equation}
\alpha \mathcal{K}_{ij} 
	\approx 
	\alpha \overline{\mathcal{K}}_{ij}
	=
	\alpha \mathscr{K}_{ij}
		\, ,
\qquad \quad
\alpha \rho^{ij} 
	\approx
	\alpha \overline{\rho}^{ij}
	= 
	0 \, ,
\label{KrhoLeading}
\end{equation}
so that the multiplier is,
\begin{equation}
\alpha \mathcal{F}_{ij}
	\approx
	\alpha \overline{\mathcal{F}}_{ij}
	=
    - \alpha \Bigl( 
        \mathscr{Q}_{ij} - \frac{ g_{ij} }{2} \mathscr{Q} 
        \Bigr)
        \, .
\label{Fleading}
\end{equation}
In the language of Dirac's formalism~\cite{DiracBook} for theories with constraints,~$\Phi^{ij}$ and~$\Psi^{ij}$
are second-class constraints whose conservation determines the Lagrange multiplier~$\mathcal{F}_{ij}$.
This is the mathematical structure stemming from the requirement of analyticity in~$\alpha$
that allows to excise the degrees of freedom associated to higher derivatives.
This produces the canonical action at leading order,
\begin{align}
\MoveEqLeft[6]
\alpha \mathscr{S}_{\rm red} \bigl[ N, N_i, g_{ij}, \pi^{ij} \bigr] 
	\equiv
\alpha \mathcal{S} \bigl[ N, N_i, g_{ij}, \overline{\mathcal{K}}_{ij}, 
	\overline{\mathcal{F}}_{ij}, \pi^{ij}, \overline{\rho}^{ij} \bigr] 
\nonumber \\
={}&
	\alpha \! \int\! d^{4\!} x \, \Bigl[
		\pi^{ij} \dot{g}_{ij}
		-
		N \mathcal{H}_{\rm red}
		-
		N_i \mathcal{H}_{\rm red}^i
		\Bigr]
\, ,
\end{align}
where the reduced Hamiltonian and momentum constraints,
\begin{equation}
\alpha \mathcal{H}_{\rm red} 
	= 
	- \alpha \sqrt{g} \, \frac{ ( \mathscr{Q} - 2\Lambda ) }{\kappa^2} 
	\, ,
\qquad \quad
\alpha \mathcal{H}_{\rm red}^i
	=
	- \alpha \sqrt{g} \, \frac{ 2 \mathscr{L}^i }{ \kappa^2}
	\, ,
\end{equation}
are written in terms of quantities given in~(\ref{LiQdef}).

%

%%%%%%%%%%%%%%%%%%%%%%%%%%%%%%%%%%%%%%%%%%
%%%%%%%%%%%%%%%%%%%%%%%%%%%%%%%%%%%%%%%%%%
%%%	EQUATIONS AT SUBLEADING ORDER
%%%%%%%%%%%%%%%%%%%%%%%%%%%%%%%%%%%%%%%%%%
%%%%%%%%%%%%%%%%%%%%%%%%%%%%%%%%%%%%%%%%%%
\subsection{Equations at subleading order}
\label{subsec: Equations at subleading order}

In order to determine the subleading order of the reduced action we need 
to consider the equations of motion~(\ref{EOM1})--(\ref{EOM7}) including 
terms of order~$\alpha$.
Given that we have solutions of algebraic equations 
at leading order in~(\ref{KrhoLeading}) and~(\ref{Fleading})
we plug in those solutions directly,
\begin{equation}
\alpha {\mathcal{Q}^{ij}}_{kl} \approx
    \alpha {\overline{\mathcal{Q}}^{ij}}_{kl}
    = \alpha {\mathscr{Q}^{ij}}_{kl} \, ,
\qquad \quad
\alpha {\mathcal{L}^i}_{jk}
    \approx \alpha {\overline{\mathcal{L}}^i}_{jk}
    = \alpha {\mathscr{L}^i}_{jk} \, .
\end{equation}
Thus, the resulting equations descending from varying with canonical 
pairs are,
\begin{align}
\frac{\delta \mathcal{S}}{\delta \mathcal{\rho}^{ij}}
	={}&
	\dot{\mathcal{K}}_{ij}
		+ N \mathcal{K}_{ik} {\mathcal{K}^k}_j
		- N^k \nabla_k \mathcal{K}_{ij}
		- 2 \mathcal{K}_{k(i} \nabla_{j)} N^k
		+ \nabla_i \nabla_j N + N \mathcal{F}_{ij}
		\approx 0 
		\, ,
\label{subEOM1}
\\
\frac{\delta \mathcal{S}}{\delta \mathcal{K}_{ij}}
	={}&
    \! -
    \dot{\rho}^{ij}
    \!+
    \sqrt{g}
    \biggl[
    2 \bigl( N {\mathcal{K}_k}^{(i}
    \!-\!
    \nabla_k N^{(i}
    \bigr)
    \frac{ \rho^{j)k} }{ \sqrt{g} } 
    +
    \nabla_k \Bigl( N^k \frac{\rho^{ij} }{ \sqrt{g} } \Bigr)
    +
    2N
    \Bigl(
    \frac{ \pi^{ij} }{ \sqrt{g} }
    \!+\!
    \frac{ \mathcal{K}^{ij} \!-\! g^{ij} \mathcal{K} }{\kappa^2} 
    \Bigr)
    \biggr]
\nonumber \\
&   - 12 \alpha \kappa^2 N \sqrt{g} \,
        \biggl[
            4 \nabla_l {\overline{\mathcal{L}}_k}^{l(i} 
                \overline{\mathcal{F}}^{j)k} 
            +
            \overline{\mathcal{Q}}^{klm(i} 
                {\overline{\mathcal{Q}}^{j)n}}_{kl} 
                \overline{\mathcal{K}}_{mn}    
\nonumber \\
&    \hspace{3cm}
    -
    2 \overline{\mathcal{L}}^{mk(i} 
        {\overline{\mathcal{L}}_m}^{j)l} 
        \overline{\mathcal{K}}_{kl}
    +
    \frac{2}{N} 
        \nabla_m \bigl( 
        N {\overline{\mathcal{Q}}_{kl}}^{m(i} 
        \overline{\mathcal{L}}^{j)kl} \bigr)
    \biggr]
    \, ,
\label{subEOM2}
\\
\frac{\delta \mathcal{S}}{\delta \mathcal{\pi}^{ij}}
	={}&
	\dot{g}_{ij} - 2 \nabla_{(i} N_{j)} + 2 N \mathcal{K}_{ij}
	\approx 0
	\, ,
\label{subEOM3}
\\
\frac{\delta \mathcal{S}}{\delta g_{ij}}
	={}&  -
    \dot{\pi}^{ij}
    +
    \frac{N \sqrt{g} }{\kappa^2}  \,
    \biggl[
        -
        2 \mathcal{K}^{k(i} {\mathcal{K}^{j)}}_{k} 
        +
        2 \mathcal{K}^{ij} \mathcal{K}
        -
        G^{ij}
        +
        \frac{ g^{ij} }{2} 
            \bigl( {\mathcal{K}^i}_j { \mathcal{K}^j}_i - \mathcal{K}^2 - 2\Lambda \bigr)
        \biggr]
\nonumber \\
&\hspace{-1.3cm}
    +
    \frac{ \sqrt{g}  }{\kappa^2}
		\Bigl( 
            \nabla^i \nabla^j N
            \!-\!
            g^{ij} \nabla^k \nabla_k N
            \Bigr)
    \!+\!
    \sqrt{g} \,
        \nabla_k \biggl[ 
        N^k \frac{ \pi^{ij} }{ \sqrt{g} }
        \!-\!
        2 \frac{ \pi^{k(i} }{ \sqrt{g} } N^{j)}
        \!+\!
	    \frac{\rho^{k(i}}{\sqrt{g}} \nabla^{j)} N
        \!-\!
        \frac{1}{2}
	    \frac{ \rho^{ij} }{ \sqrt{g} } \nabla^k N
\nonumber \\
&
\hspace{-1.3cm}
        \!-\!
    2
    \frac{ \rho^{kl} }{ \sqrt{g} } {\mathcal{K}^{(i}}_l N^{j)}
    \biggr]
    +
    \rho^{kl} \Bigl(
    2 {\mathcal{K}^{(i}}_{k} \nabla_{l} N^{j)}
    +
	N^{(i} \nabla^{j)} \mathcal{K}_{kl}
    - 
    N {\mathcal{K}^{(i}}_k {\mathcal{K}^{j)}}_l
    \Bigr)
    +
    \alpha \kappa^2 \! \sqrt{g} \,
    \overline{\mathscr{G}}^{ij}
	\, ,
\label{subEOM4}
\end{align}
where the bar on the lengthy tensor~$\mathscr{G}^{ij}$, 
given in~(\ref{GijLengthy}), means it is evaluated with the leading 
order solutions for~$\mathcal{K}_{ij}$,~$\rho^{ij}$, 
and~$\mathcal{F}_{ij}$ found in the preceding subsection.
Equations descending from varying with respect to variables appearing algebraically only are,
\begin{align}
&
\frac{\delta \mathcal{S}}{ \delta N}
	=
	\sqrt{g} \biggl[
	2 \mathcal{K}_{ij} \frac{ \pi^{ij} }{ \sqrt{g} }
	+
	\bigl( \mathcal{K}_{ik} {\mathcal{K}^k}_j \!+\! \mathcal{F}_{ij}  \bigr) \frac{ \rho^{ij} }{ \sqrt{g} } 
	+ \!
	\nabla_i \nabla_j \Bigl( \frac{ \rho^{ij} }{ \sqrt{g} } \Bigr)
\nonumber \\
&	\hspace{2cm}
	+
	\frac{2}{\kappa^2} \Bigl( {\mathcal{K}^i}_j { \mathcal{K}^j}_i \!-\! \mathcal{K}^2 \!+\! R \!-\! 2 \Lambda  \Bigr)
	+ \! 4 \alpha \kappa^2 { \overline{\mathcal{F}}^i}_j \Bigl( 2 { \overline{\mathcal{F}}^j}_k { \overline{\mathcal{F}}^k}_i
		- 3 { \overline{\mathcal{L}}_i}^{kl} { \overline{\mathcal{L}}^j}_{kl} \Bigr)
\nonumber \\
&	\hspace{2cm}
	+ \alpha \kappa^2 
    { \overline{\mathcal{Q}}^{ij}}_{kl} \Bigl( { \overline{\mathcal{Q}}^{kl}}_{mn} { \overline{\mathcal{Q}}^{mn}}_{ij} 
		- 6 { \overline{\mathcal{L}}_m}^{kl} { \overline{\mathcal{L}}^m}_{ij} \Bigr)
	\biggr]
	\equiv 
    - \mathcal{H}
	\approx 0
	\, ,
\label{subEOM5}
\\
&
\frac{\delta \mathcal{S}}{ \delta N_i }
	=
	\sqrt{g}
	\biggl[
	2 \nabla_{j} \Bigl( \frac{ \pi^{ij} }{ \sqrt{g} } \Bigr)
	+ 2 \nabla_{j} \Bigl( {\mathcal{K}^i}_{k} \frac{ \rho^{jk} }{ \sqrt{g} } \Bigr)
	- \frac{ \rho^{jk} }{ \sqrt{g} } \nabla^i \mathcal{K}_{jk}
	\biggr]
	\equiv 
	- \mathcal{H}^{i}
	\approx 0
	\, ,
\label{subEOM6}
\\
&
\frac{ \delta \mathcal{S} }{ \delta \mathcal{F}_{ij} }
	=
	N \sqrt{g} \biggl[
	\frac{ \rho^{ij} }{ \sqrt{g} }
	+
	12 \alpha \kappa^2 
    \Bigl( 2 \overline{\mathcal{F}}^{ik} {\overline{\mathcal{F}}_k}^j
		- \overline{\mathcal{L}}^{ikl} {\overline{\mathcal{L}}^j}_{kl} \Bigr)
	\biggr]
	\equiv - N \Phi^{ij}
	\approx 0
	\, .
\label{subEOM7}
\end{align}
These again represent constraints, as they do not yield solutions 
for~$N$,~$N_i$, not~$\mathcal{F}_{ij}$. Rather, they define primary 
constraints~$\mathcal{H}$,~$\mathcal{H}^i$, and~$\Phi^{ij}$. 

Just as at leading order, 
the conservation of Hamiltonian and momentum constraints does not 
generate secondary constraints, while the conservation of~$\Phi^{ij}$ does, 
\begin{equation}
\dot{\Phi}^{ij} \approx
    - 2 N \sqrt{g} \biggl[
	\frac{\pi^{ij}}{ \sqrt{g} }
	+ \frac{1}{\kappa^2} 
        \bigl( \mathcal{K}^{ij} - g^{ij} \mathcal{K} \bigr)
	\biggr]
    +
    \frac{2 \alpha N}{\kappa^2}
    \bigl( \Delta \mathcal{K}^{ij} - g^{ij} \Delta \mathcal{K} \bigr)
    \equiv
    N \Psi^{ij}
    \approx 0
    \, .
\end{equation}
The precise form of the contribution~$\Delta {\cal K}^{ij}$ above 
will turn out not to be important,
apart from the fact that it depends only on the canonical variables~$g_{ij}$
and~$\pi^{ij}$ (and their derivatives), and that the conservation of~$\Psi^{ij}$
does not generate a further constraint, but rather an expression that determines
the Lagrange multiplier,
\begin{align}
\dot{\Psi}^{ij} \approx{}&
    \kappa^2 N \sqrt{g}
    \biggl[
    - 2 \frac{ \pi^{ik} }{ \sqrt{g} } 
        \frac{ {\pi_k}^j }{ \sqrt{g} } 
    + \frac{\pi}{ \sqrt{g} } \frac{ \pi^{ij} }{ \sqrt{g} } 
    + g^{ij} \Bigl(
        \frac{ \pi^{kl} }{ \sqrt{g} } \frac{ \pi_{kl} }{ \sqrt{g} } 
    - \frac{ 1 }{2}
        \frac{ \pi }{ \sqrt{g} } \frac{ \pi }{ \sqrt{g} } 
        \Bigr)
\nonumber \\
&    + \frac{2}{\kappa^4} \Bigl( G^{ij} + \mathcal{F}^{ij} 
        - g^{ij} \mathcal{F} + g^{ij} \Lambda \Bigr)
    \biggr]
    -
    \frac{ 2 \alpha N }{ \kappa^2 }
    \bigl( \Delta \mathcal{F}^{ij} - g^{ij} \Delta \mathcal{F} \bigr)
    \approx 0 \, ,
\label{dotPsiEq}
\end{align}
where we introduced a label~$\Delta {\cal F}^{ij}$
for another contribution whose exact form will turn out not to be important,
apart from the fact that it is a function of the canonical variables~$g_{ij}$
and~$\pi^{ij}$ and their derivatives.

We can now solve Eqs.~(\ref{subEOM7})--(\ref{dotPsiEq}) for 
variables~$\mathcal{K}_{ij}$, $\rho^{ij}$, and~$\mathcal{F}_{ij}$
to subleading order,
\begin{align}
\mathcal{K}_{ij} 
	\approx{}&
	\overline{\mathcal{K}}_{ij} 
	=
    \mathscr{K}_{ij}
    +
    \alpha \Delta \mathcal{K}_{ij}
    \, ,
\label{KsubleadingSolution}
\\
\rho^{ij} 
	\approx{}&
	\overline{\rho}^{ij}
	=
	0
	-
	6 \alpha \kappa^2 \! \sqrt{g} 
		\Bigl( 
        4 \mathscr{Q}^{ik} {\mathscr{Q}_k}^j
        - 
        4 \mathscr{Q}^{ij} \mathscr{Q}
        + 
        g^{ij} \mathscr{Q}^2
	    -
        2\mathscr{L}^{ikl} { \mathscr{L}^j}_{kl} \Bigr)
	\, ,
\label{RsubleadingSolution}
\\
\mathcal{F}_{ij} 
	\approx{}&
	\overline{\mathcal{F}}_{ij} 
	=
    - \Bigl( 
        \mathscr{Q}_{ij} - \frac{ g_{ij} }{2} \mathscr{Q} 
        \Bigr)
    +
    \alpha \Delta \mathcal{F}_{ij}
    \, .
\label{FsubleadingSolution}
\end{align}
The sought reduced canonical action to subleading order is then obtained by 
plugging in these solutions
into the extended action~(\ref{ExtendedAction}),
\begin{align}
\MoveEqLeft[3]
\mathscr{S}_{\rm red} \bigl[ N, N_i, g_{ij}, \pi^{ij} \bigr] 
\equiv
\mathcal{S} \bigl[ N, N_i, g_{ij}, \overline{\mathcal{K}}_{ij}, 
    \overline{\mathcal{F}}_{ij}, \pi^{ij}, \overline{\rho}^{ij} \bigr] 
\nonumber \\
={}&
    \int\! d^{4\!} x \, \Biggl\{
    \biggl[
    \pi^{ij}
	+
    \frac{1}{2} \rho^{kl} \Bigl(
		4\delta_k^{(i} {\mathscr{K}_l}^{j)}
        - g^{ij} \mathscr{K}_{kl}
        - g_{kl} \mathscr{K}^{ij}
        - 2 \delta_k^{(i} \delta_l^{j)} \mathscr{K}
        + g_{kl} g^{ij} \mathscr{K}
		\Bigr)
    \biggr]
    \dot{g}_{ij}
\nonumber \\
&
  -
    \kappa^2 \biggl(
		\frac{ \rho_{kl} }{ \sqrt{g} } 
		- \frac{g_{kl}}{2} \frac{ \rho }{ \sqrt{g} } 
		\biggr) \dot{\pi}^{kl}
    +
	N \sqrt{g} \biggl[
        \frac{ ( \mathscr{Q} - 2 \Lambda ) }{\kappa^2}
    +
    \frac{ \overline{\rho}^{ij} }{ \sqrt{g} }
        \Bigl( \mathscr{K}_{ik} {\mathscr{K}^k}_j
 	      -
        \mathscr{Q}_{ij} 
        +
        \frac{ g_{ij} }{2} \mathscr{Q} 
        \Bigr)
\nonumber \\
&
	+
    3 \alpha \kappa^2 \mathscr{Q} \Bigl(
    4 {\mathscr{Q}^i}_j {\mathscr{Q}^j}_i 
	-
    \mathscr{Q}^2
	-
    2 {\mathscr{L}_i}^{kl} {\mathscr{L}^i}_{kl} 
    \Bigr)
	-
    4 \alpha \kappa^2 {\mathscr{Q}^i}_j 
    \Bigl(
    2 {\mathscr{Q}^j}_k {\mathscr{Q}^k}_i
	-
    3 {\mathscr{L}_i}^{kl} {\mathscr{L}^j}_{kl} 
        \Bigr)
\nonumber \\
&
	+ 
    \alpha \kappa^2 {\mathscr{Q}^{ij}}_{kl} \Bigl( 
        {\mathscr{Q}^{kl}}_{mn} {\mathscr{Q}^{mn}}_{ij} 
		- 6 {\mathscr{L}_m}^{kl} {\mathscr{L}^m}_{ij} 
    \Bigr)
 	+
    \nabla_i \nabla_j
        \Bigl( \frac{\overline{\rho}^{ij} }{ \sqrt{g} } \Bigr)
    \biggr]
\nonumber \\
&
	+ N_i \sqrt{g} \biggl[
        \frac{2 \mathscr{L}^i}{\kappa^2}
        +
        \frac{ \overline{\rho}^{jk} }{\sqrt{g}} 
        \Bigl( 2\nabla_{j} {\mathscr{K}^i}_{k}
            - \nabla^i \mathscr{K}_{jk} \Bigr)
        +
        2 {\mathscr{K}^i}_{j} \nabla_{k} \Bigl(
	    \frac{ \overline{\rho}^{jk} }{\sqrt{g}}
        \Bigr)
        \biggr]
\Biggr\} 
\, .
\label{RawReducedAction}
\end{align}
Here we see the reason why the contributions~$\Delta\mathcal{K}_{ij}$ 
and~$\Delta \mathcal{F}_{ij}$, i.e. the subleading corrections 
to~$\mathcal{K}_{ij}$ and~$\mathcal{F}_{ij}$ 
are not of particular interest, as they all cancel out to order~$\alpha$ 
in the reduced action above. The reduced canonical action we obtained above 
accomplishes the task of reducing derivative order and excising degrees
of freedom associated to higher derivatives. However, we may still
considerably simplify this action, which we do in the following section, 
before analysing some of its properties.

%%%%%%%%%%%%%%%%%%%%%%%%%%%%%%%%%%%%%%%%%%
%%%%%%%%%%%%%%%%%%%%%%%%%%%%%%%%%%%%%%%%%%
%%%%%%%%%%%%%%%%%%%%%%%%%%%%%%%%%%%%%%%%%%
%%%	SIMPLIFICATIONS AND ANALYSIS OF REDUCED CANONICAL ACTION
%%%%%%%%%%%%%%%%%%%%%%%%%%%%%%%%%%%%%%%%%%
%%%%%%%%%%%%%%%%%%%%%%%%%%%%%%%%%%%%%%%%%%
%%%%%%%%%%%%%%%%%%%%%%%%%%%%%%%%%%%%%%%%%%
\section{Simplifications and analysis of reduced canonical action}
\label{sec: Simplifications and analysis of reduced canonical action}

Having derived the reduced action~(\ref{RawReducedAction}) 
in the preceding section
here we first apply a series of steps that considerably simplify it. 
Following this, we will compute the algebra of constraints, showing that
at least to the first subleading order the constraints retain their 
first-class character, implying that the reduced theory propagates
just the massless spin-$2$ graviton degree of freedom.

\bigskip

\noindent {\bf Anti-symmetrization identities.}
In three spatial dimensions the 4-tensor~${\mathscr{Q}^{ij}}_{kl}$ that has the symmetries of the Riemann tensor is fully expressible in 
terms of its contractions,
\begin{equation}
{\mathscr{Q}^{ij}}_{kl} = 
    2 \delta^i_{[k} {\mathscr{Q}^j}_{l]}
    -
    2 \delta^j_{[k} {\mathscr{Q}^i}_{l]}
    -
    \mathscr{Q} \delta^i_{[k} \delta^j_{l]}
    \, .
\end{equation}
Using this, together with the identity given by equation (\ref{antisymmid:act})
puts the reduced canonical action in the following form,
\begin{align}
\MoveEqLeft[3]
\mathscr{S}_{\rm red}  \bigl[ N, N_i, g_{ij}, \pi^{ij} \bigr]
= \int\! d^{4\!} x \, \Biggl\{
    \biggl[
    \pi^{ij}
	+
    \frac{1}{2} \rho^{kl} \Bigl(
		4\delta_k^{(i} {\mathscr{K}_l}^{j)}
        - g^{ij} \mathscr{K}_{kl}
        - g_{kl} \mathscr{K}^{ij}
        - 2 \delta_k^{(i} \delta_l^{j)} \mathscr{K}
\nonumber \\
&
        + g_{kl} g^{ij} \mathscr{K}
		\Bigr)
    \biggr]
    \dot{g}_{ij}
  -
    \kappa^2 \biggl(
		\frac{ \rho_{kl} }{ \sqrt{g} } 
		- \frac{g_{kl}}{2} \frac{ \rho }{ \sqrt{g} } 
		\biggr) \dot{\pi}^{kl}
	+
    N \sqrt{g} \biggl[
		\frac{ ( \mathscr{Q} - 2 \Lambda ) }{\kappa^2}
    +
    \frac{ \overline{\rho}^{ij} }{ \sqrt{g} } 
    \Bigl( 
        \mathscr{K}_{ik} {\mathscr{K}^k}_j
\nonumber \\
&
	    - \mathscr{Q}_{ij} + \frac{ g_{ij} }{2} \mathscr{Q} 
        \Bigr)
	- 8\alpha \kappa^2 {\mathscr{Q}^i }_j \Bigl( 
        2 {\mathscr{Q}^j }_k {\mathscr{Q}^k }_i
        - 3 {\mathscr{L}^j}_{kl} {\mathscr{L}_i}^{kl}
	+ 6 {\mathscr{L}^j}_{ik} \mathscr{L}^k
	+ 3 \mathscr{L}^j \mathscr{L}_i
        \Bigr)
\nonumber \\
&
	+ 6 \alpha \kappa^2 \mathscr{Q} \Bigl( 
	    4 { \mathscr{Q}^i }_j { \mathscr{Q}^j }_i
	    - \mathscr{Q}^2
        - 2 {\mathscr{L}^k}_{ij} {\mathscr{L}_k}^{ij}
	    + 4 \mathscr{L}^i \mathscr{L}_i 
        \Bigr)
	+
   \nabla_i \nabla_j 
        \Bigl( \frac{ \overline{\rho}^{ij} }{\sqrt{g}} \Bigr)
	\biggr] 
\nonumber \\
&
	+ N_i \sqrt{g} \biggl[
        \frac{2 \mathscr{L}^i}{\kappa^2}
        +
        \frac{ \overline{\rho}^{jk} }{\sqrt{g}} 
        \Bigl( 2\nabla_{j} {\mathscr{K}^i}_{k}
            - \nabla^i \mathscr{K}_{jk} \Bigr)
        +
        2 {\mathscr{K}^i}_{j} \nabla_{k} \Bigl(
	    \frac{ \overline{\rho}^{jk} }{\sqrt{g}}
        \Bigr)
        \biggr]
\Biggr\} 
\, .
\label{ReducedAntisym}
\end{align}

\bigskip

\noindent {\bf Canonicalization of brackets.}
The reduced canonical action will generically not have a canonical 
symplectic term, and therefore the Poisson brackets will not be canonical. 
This is a typical situation resulting from solving the second-class constraints, 
which change Poisson brackets to Dirac brackets. We should be able to 
canonicalize the brackets to first order in~$\alpha$ by shifting the 
canonical fields. This is the case with the reduced action we obtained
in~(\ref{ReducedAntisym}), that is canonicalized by the following shift
of canonical fields,
\begin{align}
\pi^{ij} \longrightarrow{}& 
	\pi^{ij}
	-
    \frac{1}{2} \overline{\rho}^{kl} \Bigl(
		4\delta_k^{(i} {\mathscr{K}_l}^{j)}
        - g^{ij} \mathscr{K}_{kl}
        - g_{kl} \mathscr{K}^{ij}
        - 2 \delta_k^{(i} \delta_l^{j)} \mathscr{K}
        + g_{kl} g^{ij} \mathscr{K}
		\Bigr)
	\, ,
\label{piCanonicalization}
\\
g_{ij} \longrightarrow{}& 
	g_{ij} 
	-
	\kappa^2
		\biggl( \frac{ \overline{\rho}_{ij} }{\sqrt{g}} 
        - \frac{g_{ij} }{2} \frac{ \overline{\rho} }{ \sqrt{g} } \biggr)
	\, .
\label{gCanonicalization}
\end{align}
These shifts induce corresponding shifts of 
leading order Hamiltonian and momentum constraints,
\begin{align}
\sqrt{g} \, \frac{ ( \mathscr{Q} - 2\Lambda ) }{\kappa^2}
    \longrightarrow{}&
    \sqrt{g} \biggl[ 
	\frac{ ( \mathscr{Q} - 2\Lambda ) }{\kappa^2}
	+
    \frac{ (\mathscr{Q} - 2\Lambda) }{4} \frac{ \overline{\rho} }{\sqrt{g}}
	-
	\nabla^i \nabla^j \Bigl( \frac{ \overline{\rho}_{ij} }{\sqrt{g}} \Bigr)
\nonumber \\
&   \hspace{3.cm}
 	- \frac{ \overline{\rho}^{ij} }{\sqrt{g}}  
        \Bigl( \mathscr{K}_{ik} {\mathscr{K}^k}_j  
		- \mathscr{Q}_{ij} + \frac{g_{ij}}{2} \mathscr{Q}
		\Bigr) 
	\biggr] \, ,
\\
\sqrt{g} \, \frac{2 \mathscr{L}^i}{\kappa^2}
    \longrightarrow{}&
        \sqrt{g}
        \biggl[
        \frac{2\mathscr{L}^i}{\kappa^2}
        +
        2 \biggl(
		\frac{ \overline{\rho}^{ij} }{\sqrt{g}}
        - \frac{ g^{ij} }{2} \frac{ \overline{\rho} }{\sqrt{g}}
		\biggr) \mathscr{L}_j
\nonumber \\
&   \hspace{1.5cm}
    -
        \frac{ \overline{\rho}^{jk} }{\sqrt{g}} 
        \Bigl(
		2 \nabla_j {\mathscr{K}^{i}}_k
        - \nabla^i \mathscr{K}_{jk}
		\Bigr)
        -
        2
        {\mathscr{K}^i}_j
            \nabla_k \Bigl( \frac{ \overline{\rho}^{jk} }{\sqrt{g}} \Bigr)
        \biggr]
        \, .
\end{align}
The reduced canonical action resulting from the above
transformations then reads,
\begin{align}
\MoveEqLeft[1.5]
\mathscr{S}_{\rm red} \bigl[ N, N_i, g_{ij}, \pi^{ij} \bigr]
= \int\! d^{4\!} x \, \biggl\{
    \pi^{ij} \dot{g}_{ij}
	+
    N \sqrt{g} \biggl[
	\frac{\mathscr{Q} - 2\Lambda}{\kappa^2}
    +
    \frac{\mathscr{Q} - 2\Lambda}{ 4 }
         \frac{ \overline{\rho} }{\sqrt{g}}
\label{CanonicalizedReducedAction}
\\
&
    - 8\alpha \kappa^2 {\mathscr{Q}^i }_j \Bigl( 
        2 {\mathscr{Q}^j }_k {\mathscr{Q}^k }_i
        - 3 {\mathscr{L}^j}_{kl} {\mathscr{L}_i}^{kl}
	+ 6 {\mathscr{L}^j}_{ik} \mathscr{L}^k
	+ 3 \mathscr{L}^j \mathscr{L}_i
        \Bigr)
	+ 3\alpha \kappa^2 \mathscr{Q} \Bigl( 
	    8 { \mathscr{Q}^i }_j { \mathscr{Q}^j }_i
\nonumber \\
&
	- 
 2\mathscr{Q}^2
     - 4 {\mathscr{L}^k}_{ij} {\mathscr{L}_k}^{ij}
	+ 8 \mathscr{L}^i \mathscr{L}_i 
        \Bigr)
	\biggr] 
	+ N_i \sqrt{g} \biggl[
        \frac{2\mathscr{L}^i}{\kappa^2}
        +
        2 \biggl(
		\frac{ \overline{\rho}^{ij} }{\sqrt{g}}
        - \frac{ g^{ij} }{2} \frac{ \overline{\rho} }{\sqrt{g}}
		\biggr) \mathscr{L}_j
        \biggr]
\biggr\} 
\, .
\nonumber 
\end{align}

\bigskip

\noindent {\bf Shifting Lagrange multipliers.}
The canonicalized reduced action in~(\ref{CanonicalizedReducedAction})
can be further simplified by perturbative redefinitions of lapse and shift fields.
The following transformations,
\begin{align}
N \longrightarrow{}&
    N 
	-
    \frac{N\kappa^2}{4} \frac{\overline{\rho}}{\sqrt{g}}
    - 3 N \alpha \kappa^4
    \Bigl( 
	    8 { \mathscr{Q}^i }_j { \mathscr{Q}^j }_i
	    - 2\mathscr{Q}^2
     - 4 {\mathscr{L}^k}_{ij} {\mathscr{L}_k}^{ij}
	+ 8 \mathscr{L}^i \mathscr{L}_i 
        \Bigr) 
\nonumber \\
&     +
    12N\alpha \kappa^4 \Lambda(\kappa^{2}{\cal H}_{0}+4\Lambda)
    + 
    24N\alpha \kappa^4 \mathscr{L}^i \mathscr{L}_i 
	\, , \label{lapsechange}
\\
N_i \longrightarrow{}&
	N_i
	-
    \kappa^2 \biggl( \frac{\overline{\rho}_{ij}}{\sqrt{g}}
        - \frac{g_{ij}}{2} \frac{ \overline{\rho} }{ \sqrt{g} } \biggr) N^j 
    +
    12 N \alpha \kappa^4
        \Bigl( 2 {\mathscr{Q}^k}_l{\mathscr{L}^l}_{ki} 
            + {\mathscr{Q}^j}_i \mathscr{L}_j
            + \mathscr{Q} \mathscr{L}_i \Bigr)
	\, , \label{shiftchange}
\end{align}
effectively remove every term containing~$ \mathscr{L}_i$, and
substitute every fully contracted~$\mathscr{Q}$ by~$2\Lambda$,\footnote{Even 
though algorithmically one can implement these simplifications by evaluating
zeroth order equations of motion on the subleading terms, this should
not be confused with the proper origin of the simplifications which is
perturbative field redefinitions.}
thus producing the final simplified reduced canonical action,
\begin{align}
\MoveEqLeft[1.5]
\mathscr{S}_{\rm red} \bigl[ N, N_i, g_{ij}, \pi^{ij} \bigr] 
    =\int\! d^{4\!} x \, 
    \Bigl[
    \pi^{ij}
    \dot{g}_{ij}
	-
    N \mathcal{H}_{\rm red}
    -
    N_i \mathcal{H}^i_{\rm red}
    \Bigr]
\, ,
\label{reducedcanonicalaction}
\end{align}
where the reduced Hamiltonian and momentum density are, respectively,
\begin{align}
\mathcal{H}_{\rm red} ={}&
    \sqrt{g} \biggl[
		- \frac{ ( \mathscr{Q} - 2 \Lambda ) }{\kappa^2}
	+ 8\alpha \kappa^2 {\mathscr{Q}^i }_j \Bigl( 
        2 {\mathscr{Q}^j }_k {\mathscr{Q}^k }_i
        - 3 {\mathscr{L}^j}_{kl} {\mathscr{L}_i}^{kl}
        \Bigr)
\nonumber \\
&   \hspace{4cm}
	- 24\alpha \kappa^2 \Lambda \Bigl( 
	    2 { \mathscr{Q}^i }_j { \mathscr{Q}^j }_i
	    - 2 \Lambda^2
        - {\mathscr{L}_k}^{ij} {\mathscr{L}^k}_{ij}
        \Bigr)
	\biggr] 
    \, ,
\label{reducedH}
\\
\mathcal{H}_{\rm red}^i ={}&
    \sqrt{g}
        \biggl[ - \frac{2 \mathscr{L}^i}{\kappa^2} \biggr]
    \, ,
\label{reducedHi}
\end{align}
where recall that expressions for quantities~${\mathscr{Q}^{ij}}_{kl}$ 
and~${\mathscr{L}^{i}}_{jk}$ and their contractions are given 
in~(\ref{KLQdefinitions}) and~(\ref{KLQcontractions})
in terms of the fields $g_{ij}$ and $\pi^{ij}$, and their spatial derivatives.

\bigskip

\noindent {\bf Constraint algebra.}
After all the simplifications made in this section we are ready to check 
the algebra of reduced constraints. The Poisson bracket for two 
functionals~$A$ and~$B$ of phase space 
variables $h_{ij}$ and $\pi^{ij}$ takes the canonical form,
\begin{equation}
\bigl\{ A , B \bigr\} 
    \equiv \int\! d^{3}x 
    \bigg[
    \frac{\delta A}{\delta h_{ij}(t,\vec{x})}
    \frac{\delta B}{\delta \pi^{ij}(t,\vec{x})}
    -
    \frac{\delta A}{\delta \pi^{ij}(t,\vec{x})}
    \frac{\delta B}{\delta h_{ij}(t,\vec{x})}
    \bigg]
    \, ,
\end{equation}
according to the symplectic part of the canonical 
action~(\ref{reducedcanonicalaction}).
For constraint algebra, it is best to first define smeared constraints,
with respect to some scalar~$f$ and some vector~$f_i$,
\begin{align}
\mathcal{H}^{\rm red} [ f ] \equiv
    \int\! d^{3\!}x \, \mathcal{H}^{\rm red}(t,\vec{x}) f(t,\vec{x})
    \, ,
\qquad
\mathcal{H}^{\rm red}_i [ f^i ] \equiv
    \int\! d^{3\!}x \, \mathcal{H}_i^{\rm red}(t,\vec{x}) f^i(t,\vec{x})
    \, ,
\end{align}
for which Poisson brackets are more convenient to evaluate. 
The algebra of the constraints~${\cal H}^{\rm red}$ 
and~${\cal H}_{i}^{\rm red}$ to first order in~$\alpha$ evaluates to:
\begin{align}
\MoveEqLeft[1]
\bigl\{ \mathcal{H}^{\rm red} [ f ] , \mathcal{H}^{\rm red} [s] \bigr\}
    =
  \mathcal{H}^{\rm red}_i \bigl[ f \nabla^i s - s \nabla^i f \bigr]
    + 
    24 \alpha \kappa^4
    \mathcal{H}^{\rm red}_i
        \Bigl[
        {\mathscr{L}_k}^{li} {\mathscr{L}_l}^{kj}
        \bigl( f\nabla_j s \!-\! s \nabla_j f \bigr)
        \Bigr]
\nonumber \\
& 
-
    24 \alpha \kappa^4
    \mathcal{H}^{\rm red}_i
        \Bigl[
        \bigl(
        4 {\mathscr{Q}^i}_j {\mathscr{Q}^j}_k
        -
        2 ( \mathscr{Q} + \Lambda ) { \mathscr{Q}^{i} }_k
        +
        {\mathscr{L}^i}_{jk} {\mathscr{L}^j} 
        \bigr)
        \bigl( f\nabla^k s - s \nabla^k f \bigr)
        \Bigr]
\nonumber \\
&
+
    24 \alpha \kappa^4
    \mathcal{H}^{\rm red}_i
        \Bigl[
        \bigl(
        2 {\mathscr{Q}^j}_{[k} {\mathscr{Q}^k}_{j]}
        +
        2 \Lambda \mathscr{Q}
        +
        {\mathscr{L}^j}_{kl} {\mathscr{L}_j}^{kl}
        -
        \mathscr{L}^j \mathscr{L}_j
        \bigr)
        \bigl( f\nabla^i s - s \nabla^i f \bigr)
        \Bigr]
\nonumber \\
&
    +
    24 \alpha \kappa^4
    \mathcal{H}^{\rm red}
    \Bigl[ 
    \bigl( {\mathscr{Q}^i }_j {\mathscr{L}^j}_{ik} 
        - \Lambda \mathscr{L}_{k} \bigr)
    (f \nabla^{k} s - s \nabla^{k} f )
    \Bigr]
    +
    \mathcal{O}(\alpha^2)
    \approx
    0
    +
    \mathcal{O}(\alpha^2)
    \, , \label{PBHH}
\\
\MoveEqLeft[1]
\bigl\{ \mathcal{H}^{\rm red} [ f ] , \mathcal{H}^{\rm red}_i [ s^i ] \bigr\}
    =
    -
    \mathcal{H}^{\rm red} \bigl[ s^i \nabla_i f \bigr]
    \approx 0
    \, , \label{PBHHi}\\
\MoveEqLeft[1]
\bigl\{ \mathcal{H}^{\rm red}_i [ f^i ] , 
    \mathcal{H}^{\rm red}_j [ s^j ] \bigr\}
    =
    \mathcal{H}^{\rm red}_i \bigl[ f^j \nabla_j s^i
        - s^j \nabla_j f^i\bigr]
        \approx 0
        \, ,
\label{PBHiHj}
\end{align}
where in (\ref{PBHH}) we have used the antisymmetrization identities (\ref{antisymmid:pb1}) and (\ref{antisymmid:pb2}). 

Using Hamilton's equations of motion that follow from varying the reduced canonical action (\ref{reducedcanonicalaction}), the evolution of a functional~$A$ on phase space is given by:
\begin{equation}
\dot{A} =
    \bigl\{A,{\cal H}^{\rm red}[N]
    +
    {\cal H}_{i}^{\rm red}[N^{i}]
    \bigr\}
\label{dotF}
\end{equation}
It follows then that the time evolution of ${\mathcal H}^{\rm red}$ 
and $\mathcal{H}^{\rm red}_{i}$ follow from the set of 
equations~(\ref{PBHH})--(\ref{PBHiHj}). These equations imply that if the constraints hold at some initial moment of time then they hold at later times to order $\alpha$. Therefore, time evolution according to the Hamiltonian (\ref{dotF}) to order $\alpha$ does not generate any further constraints and the constraint analysis is completed.

\bigskip

\noindent{\bf Further canonical transformations.}
A question remains of whether we can absorb the correction to the Hamiltonian
constraint~(\ref{reducedH}), obtained by excising spurious degrees of freedom, 
into further field redefinitions. In other words, we still need to demonstrate
that the reduction procedure yields a theory different from general relativity
to order~$\alpha$. To this end we consider a canonical transformation of the 
spatial metric and its conjugate momentum, so that the Poisson brackets are
preserved, as they already have the same canonical form as in general relativity.
The canonical transformation takes a relatively simple form when considered
linear order in~$\alpha$ only,
\begin{equation}
g_{ij} \longrightarrow{} g_{ij}
    + 
    \alpha \frac{\delta \mathcal{G}}{ \delta \pi^{ij} } 
    \, ,
\qquad \quad
\pi^{ij} \longrightarrow{} \pi^{ij}
    - 
    \alpha \frac{\delta \mathcal{G}}{ \delta g_{ij} } 
    \, ,
\label{CanonicalTrans}
\end{equation}
where it is encoded by the generating functional,
\begin{equation}
\mathcal{G} [ g_{ij}, \pi^{ij} ]
    =
    \int\! d^{3\!}x \, \sqrt{g} \
    \mathscr{G}\Bigl[ g_{ij}, \frac{\pi^{ij}}{\sqrt{g}}, R_{ij}, \nabla_i \Bigr] \, ,
\end{equation}
that is an integral over the local spatial scalar~$\mathscr{G}$. This last 
property guarantees that the momentum constraint~(\ref{reducedHi}) is not
changed by the canonical transformation. Only the Hamiltonian constraint
transforms under such a canonical transformation,
\begin{align}
&
\mathcal{H}_{\rm red} 
    \longrightarrow
    \biggl[
    1
    -
    \frac{\alpha g^{ij}}{2} 
        \frac{\delta \mathcal{G}}{ \delta \pi^{ij} }
    \biggr]
    \mathcal{H}_{\rm red}
    +
    \frac{ \alpha \sqrt{g} }{2 \kappa^2 }
    \biggl[
    R^{ij}
    -
    g^{ij}
    ( R - 2 \Lambda )
    -
    \nabla^i \nabla^j 
    +
    g^{ij} \nabla^k \nabla_k
    \biggr]
    \frac{\delta \mathcal{G}}{ \delta \pi^{ij} }
\nonumber \\
&   \hspace{0.4cm}
    +
    \alpha \kappa^2 \sqrt{g}
        \biggl[
		\frac{ \pi^{ik} }{ \sqrt{g} }
		\frac{ {\pi_k}^{j} }{ \sqrt{g} }
		-
        \frac{1}{2}
		\frac{ \pi^{ij} }{ \sqrt{g} }
		\frac{ \pi }{ \sqrt{g} }
		\biggr]
        \frac{\delta \mathcal{G}}{ \delta \pi^{ij} } 
    -
    \alpha \kappa^2
    \biggl[ \frac{ \pi_{ij} }{ \sqrt{g} }
		-
		\frac{g_{ij}}{2}
		\frac{ \pi }{ \sqrt{g} }
		\biggr]
        \frac{\delta \mathcal{G}}{ \delta g_{ij} } 
    +
    \mathcal{O}(\alpha^2)
    \, .
\label{Htransformed}
\end{align}
The multiplicative factor in the first term can be absorbed into a
perturbative redefinition of lapse, and can thus effectively be set to one 
here. The remaining terms can potentially remove the correction to the
Hamiltonian constraint linear in~$\alpha$.

It is sufficient that at least a single term in~(\ref{reducedH}) 
of order~$\alpha$
cannot be removed by canonical transformation to demonstrate that the
reduced theory we find is not equivalent to general relativity. To this 
end it is sufficient to focus on the term in the Hamiltonian constraint
containing six factors of canonical momenta,
\begin{align}
\mathcal{H}_{\rm red}
    \supset{}&
    16 \alpha \kappa^2 \sqrt{g} \, 
    {\mathscr{Q}^i}_j {\mathscr{Q}^j}_k {\mathscr{Q}^k}_i
\nonumber \\
    \supset{}&
    2 \alpha \kappa^{14} \sqrt{g}
    \biggl[
    - 8 
    \biggl(
    \frac{{\pi^i}_j}{\sqrt{g}}
    \frac{ {\pi^j}_k }{\sqrt{g}}
    \frac{{\pi^k}_l}{\sqrt{g}}
    \frac{ {\pi^l}_m }{\sqrt{g}}
    \frac{{\pi^m}_n}{\sqrt{g}}
    \frac{ {\pi^n}_i }{\sqrt{g}}
    \biggr)
    +
    12 
    \biggl( 
    \frac{{\pi^i}_j}{\sqrt{g}} 
    \frac{ {\pi^j}_l }{\sqrt{g}}
    \frac{{\pi^l}_k}{\sqrt{g}} 
    \frac{ {\pi^k}_m }{\sqrt{g}}
    \frac{{\pi^m}_i}{\sqrt{g}} 
    \biggr)
    \frac{\pi}{\sqrt{g}}
\nonumber \\
&   \hspace{1.5cm}
    -
    6
    \biggl(
    \frac{{\pi^i}_j}{\sqrt{g}}
    \frac{ {\pi^j}_k }{\sqrt{g}}
    \frac{{\pi^k}_l}{\sqrt{g}} 
    \frac{{\pi^l}_i}{\sqrt{g}} 
    \biggr)
    \Bigl(
    \frac{\pi}{\sqrt{g}}
    \Bigr)^{\!2}
    +
    \biggl( \frac{{\pi^i}_j}{\sqrt{g}} 
    \frac{{\pi^j}_k}{\sqrt{g}} 
    \frac{{\pi^k}_i}{\sqrt{g}} \biggr)
    \Bigl(
    \frac{\pi}{\sqrt{g}}
    \Bigr)^{\!3}
    \biggr]
    \, .
\label{Hmomenta}
\end{align}
The traces over four or more factors of momenta can be rewritten in terms
of traces over three or less factors of momenta by virtue of the 
Cayley-Hamilton theorem (see e.g.~appendix B 
from~\cite{Yao:2020tur}). Using the identities collected in 
appendix~\ref{app:CH}
the terms in~(\ref{Hmomenta}) can be expressed in terms the traces over a maximum product of three matrices ${\pi^{i}}_{j}$. Among those terms there is 
a particular one we want to focus on,
\begin{equation}
\mathcal{H}_{\rm red}
    \supset
- \frac{16 }{3} \alpha \kappa^{14} 
    \sqrt{g}
    \biggl(
    \frac{{\pi^i}_j}{\sqrt{g}}
    \frac{ {\pi^j}_k }{\sqrt{g}}
    \frac{{\pi^k}_i}{\sqrt{g}}
    \biggr)
    \biggl(
    \frac{ {\pi^l}_m }{\sqrt{g}}
    \frac{{\pi^m}_n}{\sqrt{g}}
    \frac{ {\pi^n}_l }{\sqrt{g}}
    \biggr)
    \, .
\label{H33}
\end{equation}
The terms above can potentially be absorbed into terms generated by the
canonical transformation in~(\ref{Htransformed}). Such terms can descend
only from a particular part of the generating functional,
\begin{equation}
\mathcal{G}
    \supset
    c \times \kappa^{12} \!\!
    \int\! d^{3\!}x \, \sqrt{g} \,
    \biggl( \frac{ {\pi^i}_j }{ \sqrt{g} }
        \frac{ {\pi^j}_k }{ \sqrt{g} }
        \frac{ {\pi^k}_i }{ \sqrt{g} } \biggr)
    \biggl( \frac{ {\pi^l}_m }{ \sqrt{g} }
        \frac{ {\pi^m}_l }{ \sqrt{g} } \biggr)
    \, .
\end{equation}
Plugging this into the expression~(\ref{Htransformed}) we find that no
terms of the form~(\ref{H33}) are generated by the canonical transformation.
Therefore, the reduced theory that we find is not equivalent to general 
relativity in the absence of matter. Rather, it belongs to the Type-II class of minimally modified
gravity theories~\cite{Aoki:2018brq}.

%%%%%%%%%%%%%%%%%%%%%%%%%%%%%%%%%%%%%%%%%%
%%%%%%%%%%%%%%%%%%%%%%%%%%%%%%%%%%%%%%%%%%
%%%%%%%%%%%%%%%%%%%%%%%%%%%%%%%%%%%%%%%%%%
%%%	DISCUSSION
%%%%%%%%%%%%%%%%%%%%%%%%%%%%%%%%%%%%%%%%%%
%%%%%%%%%%%%%%%%%%%%%%%%%%%%%%%%%%%%%%%%%%
%%%%%%%%%%%%%%%%%%%%%%%%%%%%%%%%%%%%%%%%%%
\section{Discussion}
\label{sec: Discussion}

Local effective field theories are a useful tool for studying effects of  
unknown ultraviolet sectors of physical theories on low energy physics. 
They are meant to capture these 
corrections solely in terms of low energy degrees of freedom by introducing
effective interaction terms into the action, as dictated by the assumed
symmetries. However, such local effective theories are generically
restricted to yield solutions solely in the form of a perturbative series.
This is because they suffer from an
unpleasant feature when considered at face value: some effective interaction
terms will generically contain higher derivatives, which introduces
additional propagating degrees of freedom not present in the low energy 
theory. 
This feature is an artifact of demanding locality of the effective theory. 
In order to understand and map out 
the consequences of some yet unknown ultraviolet physics 
in effective theories of gravity,
the truncated equations of motion need to be treated carefully, so that 
predictions do not depend on additional degrees of freedom introduced by the
truncation.
For this purpose it is useful to find a way to explicitly excise the 
spurious degrees of freedom associated to higher derivative terms.

In the present paper we have applied an action-based derivative reduction 
procedure to an effective theory of gravity in~(\ref{action}), 
consisting of the leading Einstein-Hilbert term with a cosmological constant, 
and supplemented by a Riemann-cubed term with a dimensionless 
coupling constant~$\alpha$. The reason for considering this
particular EFT is that all the curvature-squared terms, and all other 
curvature-cubed terms can be removed from the action by disformal field
redefinitions of the metric, which makes this the simplest non-trivial case.
Our reduction procedure was performed in the canonical (Hamiltonian)
formalism where the number of propagating degrees of freedom is manifest.
The usual algorithm for deriving the canonical formalism, starting from the 
Lagrangian one, is supplemented by the requirement that all the dynamical
quantities are formally analytic functions of~$\alpha$. This requirement
fashions effective second-class constraints that eliminate the would-be
degrees of freedom associated to higher derivatives.

While the derivative reduction procedure we consider is known to work for 
theories without constraints~\cite{Glavan:2017srd,Jaen:1986iz,Eliezer:1989cr}, 
it was an open question whether the procedure 
interferes with the constraint structure of gauge theories. Our results
show that the theory in~(\ref{reducedcanonicalaction}) resulting from the 
reduction procedure retains the ADM structure of general relativity,
where the momentum constraint~(\ref{reducedHi}) is that of general 
relativity, while the Hamiltonian constraint~(\ref{reducedH})
receives additional contributions that cannot be absorbed into field 
redefinitions. We paid great attention to working out the constraint 
algebra of the reduced theory, and showed that to the order of~$\alpha$
we are working the constraints are first-class, so that their Poisson brackets vanish on-shell. Therefore, to the order 
in $\alpha$ that we work, the reduced theory has the same number of degrees 
of freedom as general relativity and smoothly reduces to it
in the limit~$\alpha\!\to\!0$. It should be noted that the reduction 
procedure is in essence perturbative, but that it can in principle be 
extended to an arbitrary order.

Even though the reduced theory preserves the constraint algebra of general
relativity on-shell, it does not preserve it off-shell. This implies the
physical interpretation of constraints in the reduced theory differs
from that of general relativity, despite all constraints remaining first-class.
In retrospect, this feature is not surprising, as the reduction procedure
effectively selects a preferred frame at short distances by trading higher time derivatives 
with spatial derivatives and potential terms. Thus, spatial diffeomorphisms
are preserved in the same form as in general relativity, but time 
diffeomorphisms are deformed into a different local symmetry. This is 
precisely the structure of {\it minimally modified 
gravity}~\cite{Lin:2017oow,Mukohyama:2019unx}, that was introduced
previously by classical considerations of constructing modified gravity 
theories. Our reduced theory cannot be brought to the
general relativity form upon field redefinitions, implying it is 
indeed distinct and falls under type-II class of 
minimally modified gravity~\cite{Aoki:2018brq}.

There are a number of other interesting questions to be addressed in future 
works. We have presented the derivative reduction procedure to 
order~$\mathcal{O}(\alpha)$. 
Even though it seems reasonable to expect the procedure to work for 
higher orders in~$\alpha$, demonstrating this remains an open question. 
For example, a successful application of 
the order reduction algorithm to $\mathcal{O}(\alpha^{2})$ would involve an 
extension of \eqref{reducedcanonicalaction} such that the 
constraints~(\ref{reducedH}) and~(\ref{reducedHi}) would now contain terms 
of~$\mathcal{O}(\alpha^{2})$ and with all Poisson brackets between the set 
of constraints weakly vanishing to $\mathcal{O}(\alpha^{2})$. This task should 
be straightforward, though mathematically more involved. The intriguing
question is whether the reduction procedure can be extended to all orders,
thus obtaining a theory in which the effects of higher derivatives are
fully taken into account, with the spurious degrees of freedom completely
removed. We have no strong reason to argue either in favour or against,
though the existence of minimally modified gravity theories might hint
toward this being possible, at least in some capacity. 
One part of addressing this question will have to be understanding better
the role of perturbative field redefinitions~\cite{Daas:2024pxs}
that simplify the reduced theory.
It is particularly
interesting to ponder whether there is a perturbative canonical field 
transformation of the action~(\ref{reducedcanonicalaction}) that leads to
Hamiltonian and momentum constraints that have exactly vanishing Poisson 
brackets on shell, not just to leading order in~$\alpha$.

In pure gravity one can appeal to perturbative field
redefinitions of the metric in order to completely absorb quadratic 
curvature corrections, and most of the cubic corrections, allowing us
to consider the action in~(\ref{action}) as the first nontrivial
example. However, these redefinitions are not as inconsequential 
when matter is considered, as they modify the coupling of matter to gravity.
There are two possibilities that we can think of to couple matter consistently
in the reduced theory.
One is to add matter fields in the covariant action before order reduction and then to perform order reduction for the total action. It can then be checked 
after order reduction whether the constraints~${\mathcal H}_{\rm red}$ 
and ${\mathcal H}_{\rm red}^{i}$ still weakly commute to a given order in
coupling constants of higher derivative terms, confirming that no spurious
degrees of freedom are present. Another possibility is to add a 
gauge-fixing condition as an additional second-class constraint
that has non-vanishing Poisson bracket with the Hamiltonian 
constraint~\cite{Aoki:2018zcv,DeFelice:2020eju,Aoki:2020lig}, after order 
reduction but before adding matter. Such an approach hinges on the
assumption that the reduction procedure to all orders is in principle
possible, and is considered in Appendix~\ref{sec:gravitational_waves}.
There we considered linear gravitational waves propagating in cosmological 
spaces to find that, up to field redefinitions, they are the same as in
general relativity. However, considering matter coupling might break this 
degeneracy. It would be interesting to investigate whether
the reduced theory shares the features found for black hole 
solutions~\cite{Daas:2023axu,Giacchini:2024exc}
in theories containing cubic curvature terms such 
as the one in~(\ref{action}). It has also been suggested that when looking
for black hole solutions in semiclassical gravity a reduction of spatial derivative order might 
be called for~\cite{Arrechea:2022dvy,Arrechea:2023oax}, which is not connected
to the initial value problem and removal of spurious degrees of freedom
we considered here. The EFT corrections might also be relevant for 
gravitational wave studies~\cite{Endlich:2017tqa,Sennett:2019bpc}.

Though we have concentrated on gravitational theories, the algorithm can also 
be applied to actions involving higher derivatives of fields other than the 
metric tensor. A set of models that have attracted a great deal of interest are 
Horndeski scalar tensor theories \cite{Horndeski:1974wa} which are the most 
general scalar tensor theories producing covariant field equations at most second order 
in time derivatives. The order reduction algorithm allows one to consider a  
substantially wider set of models which formally have higher order time 
derivatives in the field equations, but for which the order of these higher 
time derivatives can be perturbatively reduced. 
It should be noted that such models have to be understood as 
effective local theories capturing ultraviolet corrections, and not as 
a fundamental theory. The phenomenology of such models is yet to be explored.

%%%%%%%%%%%%%%%%%%%%%%%%%%%%%%%%%%%%%%%%%%
%%%%%%%%%%%%%%%%%%%%%%%%%%%%%%%%%%%%%%%%%%
%%%	ACKNOWLEDGEMENTS
%%%%%%%%%%%%%%%%%%%%%%%%%%%%%%%%%%%%%%%%%%
%%%%%%%%%%%%%%%%%%%%%%%%%%%%%%%%%%%%%%%%%%
\section*{Acknowledgements}

We are grateful to Ruggero Noris for a detailed reading of our paper and sending
us his comments.
DG was supported by the European Union and the Czech Ministry of Education, 
Youth and Sports (Project: 
MSCA Fellowship CZ FZU I --- CZ.02.01.01/00/22\textunderscore010/0002906). SM was supported in part by Japan Society for the Promotion of Science (JSPS) Grants-in-Aid for Scientific Research No.~24K07017 and the World Premier International Research Center Initiative (WPI), MEXT, Japan. 
TZ is supported by the European Union and Polish National Science Centre (project No. 2021/43/P/ST2/02141 under the Marie Sk\l odowska-Curie grant agreement
No. 94533).

%%%%%%%%%%%%%%%%%%%%%%%%%%%%%%%%%%%%%%%%%%
%%%%%%%%%%%%%%%%%%%%%%%%%%%%%%%%%%%%%%%%%%
%%%	APPENDICES
%%%%%%%%%%%%%%%%%%%%%%%%%%%%%%%%%%%%%%%%%%
%%%%%%%%%%%%%%%%%%%%%%%%%%%%%%%%%%%%%%%%%%
\appendix

%%%%%%%%%%%%%%%%%%%%%%%%%%%%%%%%%%%%%%%%%%
%%%	ADDITIONAL MATHEMATICAL DETAILS
%%%%%%%%%%%%%%%%%%%%%%%%%%%%%%%%%%%%%%%%%%
\section{Additional expressions mathematical details}
\label{sec: Additional mathematical details}

\subsection{Lengthy expressions}

The tensor appearing in the equation of motion descending from varying
the extended action with respect to~$g_{ij}$ is
\begin{align}
&
\mathscr{G}^{ij}
= 
    24 N \sqrt{g} \,
    \biggl[
        \mathcal{F}^{m(i}
        {\mathcal{L}^{j)}}_{kl} {\mathcal{L}_m}^{kl}
    -
    \mathcal{F}_{kl} \mathcal{F}^{k(i} \mathcal{F}^{j)l}
    -
        \mathcal{F}^{kl}
        {\mathcal{L}_{km}}^{(i} {\mathcal{L}_l}^{j)m}
            -
        \frac{1}{2} {\mathcal{L}^a}_{mn} {\mathcal{L}_{al}}^{(j} {\mathcal{Q}^{i)lmn}} 
\nonumber \\
&
    +
    3
     N \sqrt{g} \,
    \biggl[
            2{\mathcal{Q}^{klmn}}
            {\mathcal{L}^{(i}}_{kl} {\mathcal{L}^{j)}}_{mn}
    +
    \Bigl(
    {\mathcal{Q}^{kl}}_{mn} {\mathcal{Q}^{mna(i}} 
    -
    2 {\mathcal{L}_m}^{kl} {\mathcal{L}}^{ma(i}  
    \Bigr)
    \Bigl(
        2 {\mathcal{K}^{j)}}_{k} \mathcal{K}_{la}
        +
        {Q^{j)}}_{akl}
        \Bigr)
    \biggr]
\nonumber \\
&
    +
     N \sqrt{g} \,
    g^{ij} \biggl[
    2 {\mathcal{F}^m}_n  \Bigl(
    2 {\mathcal{F}^n}_k {\mathcal{F}^k}_m
    \!-\! 
    3 {\mathcal{L}_m}^{kl} {\mathcal{L}^n}_{kl} 
    \Bigr)
    +
    \frac{1}{2} 
    {\mathcal{Q}^{mn}}_{kl}  
    \Bigl(
    {\mathcal{Q}^{kl}}_{ab} {\mathcal{Q}^{ab}}_{mn}
    \!-\! 
    6 {\mathcal{L}_a}^{kl} {\mathcal{L}^a}_{mn}
    \Bigr)
    \biggr]
\nonumber \\
&
    +
    12 
	\sqrt{g} \,
            \nabla_k \biggl[
            N {\mathcal{Q}^{kl}}_{ab}   
            { \mathcal{K}_l }^{(i} {\mathcal{L}}^{j)ab}
            \!+\!
            N {\mathcal{K}_l}^{(i} {\mathcal{Q}^{j)l}}_{ab}   
            {\mathcal{L}}^{kab}
            \!+\!
            N {\mathcal{Q}^{l(i}}_{ab}   
            {\mathcal{L}}^{j)ab}
            {\mathcal{K}^k}_l
            \!+\!
            2 N \mathcal{F}^{m(i} {\mathcal{K}^{j)}}_l
            {\mathcal{L}_m}^{kl}
\nonumber \\
&   \hspace{0cm}
            -
            2 N {\mathcal{L}_{l}}^{m(i} 
                {\mathcal{K}^{j)}}_{m} \mathcal{F}^{kl}
            + 
            2 N {\mathcal{L}_l}^{m(i}
                    \mathcal{F}^{j)l} {\mathcal{K}^k}_{m}
        +
        \frac{1}{2} \nabla_l 
    \Bigl(
        N \mathcal{Q}^{mnk(i} {\mathcal{Q}^{j)l}}_{mn} 
    - 
    2 N {\mathcal{L}}^{ml(i} {\mathcal{L}_m}^{j)k}
    \Bigr)
    \biggr]
    \, .
\label{GijLengthy}
\end{align}
%

%%%%%%%%%%%%%%%%%%%%%%%%%%%%%%%%%%%%%%%%%%
%%%	ANTI-SYMMETRIZATION IDENTITIES
%%%%%%%%%%%%%%%%%%%%%%%%%%%%%%%%%%%%%%%%%%
\subsection{Anti-symmetrization identities}
\label{sec: Anti-symmetrization identities}

In four spacetime dimensions any spatial tensor anti-symmetrized over four
indices vanishes identically. This fact yields several identities  useful for 
simplifying expressions.
The following identity is used to cast the canonical action into the form (\ref{ReducedAntisym}):
\begin{align}
0 = {\mathscr{Q}^i}_{[i} {\mathscr{L}_j}^{jk} {\mathscr{L}^l}_{kl]}
    ={}&
    - \frac{1}{6} {\mathscr{Q}^i}_j {\mathscr{L}_k}^{lj} {\mathscr{L}^k}_{li}
	- \frac{1}{12} {\mathscr{Q}^i}_j {\mathscr{L}^j}_{kl} {\mathscr{L}_i}^{kl}
	+ \frac{1}{3} {\mathscr{Q}^i}_j {\mathscr{L}^j}_{ik} \mathscr{L}^k
\nonumber \\
&
	+ \frac{1}{6} {\mathscr{Q}^i}_j \mathscr{L}^j \mathscr{L}_i
	+ \frac{1}{12} \mathscr{Q} {\mathscr{L}^k}_{ij} {\mathscr{L}_k}^{ij}
	- \frac{1}{6} \mathscr{Q} \mathscr{L}^i \mathscr{L}_i 
    \, , \label{antisymmid:act}
\end{align}
Additionally, the following identities are useful in simplifying the Poisson bracket (\ref{PBHH}):
\begin{align}
\MoveEqLeft[5]
0 = {\mathscr{Q}^i}_{[i} {\mathscr{Q}^j}_{j} {\mathscr{L}^{k}}_{kl]}
    =
    \frac{1}{6} {\mathscr{Q}^i}_j {\mathscr{Q}^j}_k {\mathscr{L}^k}_{il}
    +
    \frac{1}{6} {\mathscr{Q}^i}_j {\mathscr{Q}^k}_l {\mathscr{L}^j}_{ik}
    -
    \frac{1}{6} \mathscr{Q} {\mathscr{Q}^i}_j {\mathscr{L}^j}_{il}
\nonumber \\
&
    -
    \frac{1}{12} {\mathscr{Q}^i}_j {\mathscr{Q}^j}_i \mathscr{L}_l
    +
    \frac{1}{6} {\mathscr{Q}^i}_j {\mathscr{Q}^j}_l \mathscr{L}_i
    -
    \frac{1}{6} \mathscr{Q} {\mathscr{Q}^i}_l \mathscr{L}_i
    +
    \frac{1}{12} \mathscr{Q}^2 \mathscr{L}_l
    \, . 
\label{antisymmid:pb1}
\\ 
\MoveEqLeft[5]
0 = {\mathscr{L}^i}_{[i j} {\mathscr{L}_{k}}^{jl} {\mathscr{L}^k}_{m]l}
    =
    \frac{1}{12} {\mathscr{L}^i}_{kl} {\mathscr{L}_j}^{kl} {\mathscr{L}^j}_{im}
    -
    \frac{1}{12} {\mathscr{L}^i} {\mathscr{L}^j}_{ki} {\mathscr{L}^k}_{jm}
    -
    \frac{1}{12} {\mathscr{L}_m} {\mathscr{L}^i}_{jk} {\mathscr{L}_i}^{jk}
\nonumber \\
&
    +
    \frac{1}{12} \mathscr{L}_i {\mathscr{L}^i}_{jk} {\mathscr{L}_m}^{jk}
    +
    \frac{1}{12} \mathscr{L}_i \mathscr{L}^j {\mathscr{L}^i}_{jm}
    +
    \frac{1}{12} \mathscr{L}^i \mathscr{L}_i \mathscr{L}_m
    \, , 
\label{antisymmid:pb2}
\end{align}
%

%%%%%%%%%%%%%%%%%%%%%%%%%%%%%%%%%%%%%%%%%%
%%%%%%%%%%%%%%%%%%%%%%%%%%%%%%%%%%%%%%%%%%
%%% CAYLEY-HAMILTON
%%%%%%%%%%%%%%%%%%%%%%%%%%%%%%%%%%%%%%%%%%
%%%%%%%%%%%%%%%%%%%%%%%%%%%%%%%%%%%%%%%%%%
\subsection{Cayley-Hamilton theorem and trace identities}
\label{app:CH}

The three-dimensional version of the Cayley-Hamilton theorem states 
that for any three-dimensional square matrix~$\boldsymbol{A}$ the 
following relation holds,
\begin{align}
0 ={}& \boldsymbol{A}^3
    -
    {\rm Tr}(\boldsymbol{A}) \boldsymbol{A}^2
    -
    \frac{1}{2} \, {\rm Tr}(\boldsymbol{A}^2) \boldsymbol{A}
    +
    \frac{1}{2} \bigl[ {\rm Tr}(\boldsymbol{A}) \bigr]^2 \boldsymbol{A} 
\nonumber \\
&
    -
    \frac{1}{6} 
    \Bigl(
    2 \, {\rm Tr}(\boldsymbol{A}^3)
    -
    3 \, {\rm Tr}(\boldsymbol{A}^2) \, {\rm Tr}(\boldsymbol{A})
    + 
    \bigl[ {\rm Tr} (\boldsymbol{A}) \bigr]^3
    \Bigr)
    \boldsymbol{I}_3
    \, .
\end{align}
It implies that traces over four or more powers of~$\boldsymbol{A}$ 
can be written out in terms of traces over three or less powers 
of~$\boldsymbol{A}$. The three relations of interest to us are
\begin{align}
{\rm Tr}( \boldsymbol{A}^4)
    ={}&
    \frac{4}{3} \, {\rm Tr}(\boldsymbol{A}^3) \, {\rm Tr}(\boldsymbol{A})
    +
    \frac{1}{2} \bigl[ {\rm Tr}(\boldsymbol{A}^2) \bigr]^2
    - 
    {\rm Tr} ( \boldsymbol{A}^2 )
        \bigl[ {\rm Tr}(\boldsymbol{A}) \bigr]^2
    +
    \frac{1}{6} \bigl[ {\rm Tr}(\boldsymbol{A}) \bigr]^4
    \, ,
\\
{\rm Tr}(\boldsymbol{A}^5) 
    ={}&
    \frac{5}{6} \, {\rm Tr}(\boldsymbol{A}^3) \,
        {\rm Tr}(\boldsymbol{A}^2)
    +
    \frac{5}{6} \, {\rm Tr}(\boldsymbol{A}^3) 
        \bigl[ {\rm Tr}(\boldsymbol{A}) \bigr]^2
    -
    \frac{5}{6} \, {\rm Tr}(\boldsymbol{A}^2)
        \bigl[ {\rm Tr} (\boldsymbol{A}) \bigr]^3
    +
    \frac{1}{6} \bigl[ {\rm Tr}(\boldsymbol{A}) \bigr]^5
    \, ,
\\
{\rm Tr}(\boldsymbol{A}^6) 
    ={}&
    \frac{1}{3} \bigl[ {\rm Tr}(\boldsymbol{A}^3) \bigr]^2
    +
    {\rm Tr}(\boldsymbol{A}^3)
        {\rm Tr}(\boldsymbol{A}^2) \, {\rm Tr}(\boldsymbol{A})
    +
    \frac{1}{3} \, {\rm Tr}(\boldsymbol{A}^3) 
        \bigl[ {\rm Tr}(\boldsymbol{A}) \bigr]^3
    +
    \frac{1}{4} \bigl[ {\rm Tr}(\boldsymbol{A}^2) \bigr]^3
\nonumber \\
&
    - 
    \frac{3}{4}
    \bigl[ {\rm Tr}(\boldsymbol{A}^2) \bigr]^2
        \bigl[ {\rm Tr}(\boldsymbol{A}) \bigr]^2
    -
    \frac{1}{4} \, {\rm Tr}(\boldsymbol{A}^2)
        \bigl[ {\rm Tr} (\boldsymbol{A}) \bigr]^4
    +
    \frac{1}{12} \bigl[ {\rm Tr}(\boldsymbol{A}) \bigr]^6
    \, .
\end{align}
Applying these trace identities we rewrite the part of the Hamiltonian
constraint in~(\ref{Hmomenta}) as
\begin{align}
&
\mathcal{H}_{\rm red}
    \supset
    2 \alpha \kappa^{14} \sqrt{g}
    \biggl[
    -
    \frac{8}{3}
    \biggl( \frac{{\pi^i}_j}{\sqrt{g}} 
        \frac{{\pi^j}_k}{\sqrt{g}} 
        \frac{{\pi^k}_i}{\sqrt{g}} \biggr)\!
    \biggl( \frac{{\pi^l}_m}{\sqrt{g}} 
        \frac{{\pi^m}_n}{\sqrt{g}} 
        \frac{{\pi^n}_l}{\sqrt{g}} \biggr)
    \!+
    2
    \biggl( \frac{{\pi^i}_j}{\sqrt{g}} 
        \frac{{\pi^j}_k}{\sqrt{g}} 
        \frac{{\pi^k}_i}{\sqrt{g}} \biggr)\!
    \biggl( \frac{{\pi^l}_m}{\sqrt{g}} 
        \frac{{\pi^m}_l}{\sqrt{g}} \biggr)
    \frac{\pi}{\sqrt{g}}
\nonumber \\
&	\hspace{0.9cm}
    +
    \frac{1}{3} 
    \biggl( \frac{{\pi^i}_j}{\sqrt{g}} 
        \frac{{\pi^j}_k}{\sqrt{g}} 
        \frac{{\pi^k}_i}{\sqrt{g}} \biggr)
   \Bigl(  \frac{\pi}{\sqrt{g}}\Bigr)^{\!3}
    -
    2
    \biggl( \frac{{\pi^i}_j}{\sqrt{g}} 
        \frac{{\pi^j}_i}{\sqrt{g}} \biggr)
    \biggl( \frac{{\pi^k}_l}{\sqrt{g}} 
        \frac{{\pi^l}_k}{\sqrt{g}} \biggr)
    \biggl( \frac{{\pi^m}_n}{\sqrt{g}} 
        \frac{{\pi^n}_m}{\sqrt{g}} \biggr)
\nonumber \\
&	\hspace{0.9cm}
    + 
    3
    \biggl( \frac{{\pi^i}_j}{\sqrt{g}} 
        \frac{{\pi^j}_i}{\sqrt{g}} \biggr)
    \biggl( \frac{{\pi^k}_l}{\sqrt{g}} 
        \frac{{\pi^l}_k}{\sqrt{g}} \biggr)
     \Bigl(  \frac{\pi}{\sqrt{g}}\Bigr)^{\!2}
    -
    2
    \biggl( \frac{{\pi^i}_j}{\sqrt{g}} 
    \frac{{\pi^j}_i}{\sqrt{g}} \biggr)
     \Bigl(  \frac{\pi}{\sqrt{g}}\Bigr)^{\!4}
    +
    \frac{1}{3} 
     \Bigl(  \frac{\pi}{\sqrt{g}}\Bigr)^{\!6} \,
    \biggr]
    \, .
\label{HredCayleyHamilton}
\end{align}
%

%%%%%%%%%%%%%%%%%%%%%%%%%%%%%%%%%%%%%%%%%%
%%%  GRAVITATIONAL WAVES
%%%%%%%%%%%%%%%%%%%%%%%%%%%%%%%%%%%%%%%%%%
\section{Gravitational waves}
\label{sec:gravitational_waves}

In this appendix we consider gravitational waves around a flat 
Friedmann-Lema\^{i}tre-Robertson-Walker (FLRW) background. 
For this purpose we consider the following total canonical action
\begin{align}
\mathscr{S}_{\rm tot}
    =
    \mathscr{S}_{\rm red} 
    + \mathscr{S}_{\rm gaugefix} 
    + \mathscr{S}_{\rm matter} \, , 
\label{eqn:totalaction}
\end{align}
where $\mathscr{S}_{\rm red}$ is the gravitational reduced action \eqref{reducedcanonicalaction}, 
\begin{align}
\mathscr{S}_{\rm gaugefix} 
	= 
	\int\! d^{4\!}x \, N \pi \nabla^2\lambda\,,
\end{align}
is the ``gauge-fixing'' term with $\lambda$ being a Lagrange multiplier, and 
\begin{align}
\mathscr{S}_{\rm matter} 
	=
	\int\! d^{4\!} x \,  N\sqrt{g} \,P(X) \, , 
\qquad 
X = \frac{1}{N^2} \bigl( \dot{\chi} \!-\! N^i\nabla_i\chi \bigr)^2
	- \nabla^i\chi\nabla_i\chi\,,
\end{align}
is the matter action, with~$\chi$ being a k-essence matter 
field~\footnote{Recall that the lapse function $N$, shift vector $N_{i}$, and spatial 
metric $g_{ij}$ are related to the corresponding fields introduced in 
equation~(\ref{ADMmetric1}) via the transformations~(\ref{gCanonicalization}),
(\ref{lapsechange}), and (\ref{shiftchange}).}. 
We then consider the following configuration:
\begin{align}
& 
	N = \overline{N}(t) e^{\phi(t,\vec{x})} \, , 
\qquad 
	N_i = \overline{N}(t) a(t) B_i(t,\vec{x}) \, ,
\qquad
	g_{ij} = a^2(t) \bigl( e^{h(t,\vec{x})} \bigr)_{ij} \,, 
\nonumber \\
&
	\frac{\pi^{ij}}{\sqrt{g}} = \frac{ \overline{p}(t)\delta^{ij} + p^{ij}(t,\vec{x}) }{a^2(t)}
	\, , 
\qquad
	\lambda = \lambda_1(t,\vec{x}) \, , 
\qquad
	\chi = \overline{\chi}(t) + \chi_1(t,\vec{x}) \, , 
\label{eqn:FLRW+perturbation}
\end{align}
where $\phi$, $B_i$, $h_{ij}$, $p^{ij}$, $\lambda_1$ and $\chi_1$ are perturbations of 
order $\mathcal{O}(\epsilon)$, with $\epsilon$ being a small bookkeeping parameter.

By substituting \eqref{eqn:FLRW+perturbation} into the total action \eqref{eqn:totalaction} 
and expanding up to $\mathcal{O}(\epsilon)$, we obtain the following set of background 
equations.
\begin{subequations}
\begin{align}
 \frac{\partial_t \overline{\rho} }{ \overline{N} }
	+
	3H \bigl( \overline{\rho} + \overline{P} \bigr) & = 0
 \, ,
\\
 \frac{3}{4} \overline{p}^2 
 	- \Lambda
 	- \frac{\kappa^2}{2} \overline{\rho}
 	- 3\alpha\kappa^4 \bigl( \overline{p}^6 
	 	- 6\Lambda\overline{p}^4+8\Lambda^3 \bigr)
 	& = 0 \, ,
\\
\frac{\partial_t \overline{p}}{ \overline{N} }
	- \frac{\kappa^2}{2} \bigl( \overline{\rho} + \overline{P} \bigr)
    & = 0 \, ,
\\
\overline{p} 
	+ 2H 
	- 12\alpha\kappa^4 \overline{p}^{3} 
    		\bigl( \overline{p}^2-4\Lambda \bigr) & = 0\,,
\end{align}
\end{subequations}
where
\begin{align}
 H \equiv \frac{\partial_ta}{ \overline{N}a} \, ,
 \qquad 
 	\overline{\rho} \equiv 2 \overline{P}_X \overline{X} - \overline{P} \, ,
 \qquad 
 	\overline{P} \equiv P(\overline{X}) \, ,
\qquad
	\overline{P}_X \equiv P'(\overline{X}) \, . 
\end{align}
It is straightforward to show that there is no dynamical degree of freedom in the vector sector and that there is only one propagating local degree of freedom in the scalar sector as should be.

We then consider tensor perturbations by first setting
\begin{align}
\phi = 0 \, ,
\qquad 
B_i = 0 \, , 
\qquad 
h_{ij} = h^{\scr TT}_{ij} \, , 
\qquad 
p^{ij} = p_{\scr TT}^{ij} \, ,
\qquad
\lambda_1 =0 \, ,
\qquad
\chi_1 = 0 \, ,
\end{align}
where the two tensor quantities are 
traceless,~$\delta^{ij} h_{ij}^{\scr TT} \!=\! 0 \!=\! \delta_{ij} p^{ij}_{\scr TT}$,
and transverse $\delta^{jk} \partial_k h_{ij}^{\scr TT} \!=\! 0 \!=\! \partial_j p^{ij}_{\scr TT}$, 
and then expanding the total action \eqref{eqn:totalaction} up to $\mathcal{O}(\epsilon^2)$. 
The quadratic action does not contain time derivatives of $p_{\scr TT}^{ij}$ and thus, performing the Fourier transformation with respect to the spatial coordinates $\vec{x}$, the equation of motion for $p_{\scr TT}^{ij}$ can be solved algebraically. After substituting the solution for $p_{\scr TT}^{ij}$ back to the quadratic action for each Fourier component and polarization, we obtain
\begin{align}
\mathscr{S}_{\rm tensor} 
    =
    \frac{1}{4\kappa^2}
    \int\! dt \, 
    \overline{N}a^3 q_{\scr T}
    \left[ \frac{ \dot{h}_{\scr TT}^2}{ \overline{N}^2 }
        - c_{\scr T}^2\frac{k^2}{a^2}h_{\scr TT}^2\right]
        \,,
\end{align}
where
\begin{subequations}
\begin{align}
 q_{\scr T} 
 	& = 
 	1
 	+ 6 \alpha\kappa^4 
 	\left[
 	\overline{p}^2 \bigl( \overline{p}^2 - 6\Lambda \bigr)
 	+ 4 \bigl( \overline{p}^2 - 2\Lambda \bigr) \frac{k^2}{a^2}
 	\right]
 	+ \mathcal{O}(\alpha^2) 
 	\, ,
\\
 c_{\scr T}^2
 	&= 
 	1
 	+ 12 \alpha \kappa^4
 	\left[ 
 	2 \overline{p}^2 \bigl( 4\Lambda - \overline{p}^2 \bigr)
 	+ \bigl( 3 \overline{p}^2 - 2\Lambda \bigr) \frac{ \dot{ \overline{p} } }{ \overline{N} }
 	\right]
 	+ \mathcal{O}(\alpha^2)\,.
\end{align}
\end{subequations}
Here $k\!=\! \|\vec{k}\|$ denotes the modulus of the comoving wavevector $\vec{k}$.
In vacuum, i.e.~for $\overline{\rho} \!=\! \overline{P} \!=\! 0$, by using the background 
equations of motion, we obtain
\begin{equation}
 q_{\scr T} 
 	= 
 	1 
 	- 48\alpha\kappa^4 H^2
 		\Bigl( 7 H^2 \!+\! \frac{k^2}{a^2} \Bigr) 
 		+ \mathcal{O}(\alpha^2)
	\, , 
\qquad
 c_{\scr T}^2 
 	= 
 	1 
 	+ 768 \alpha\kappa^4 H^4 
 	+ \mathcal{O}(\alpha^2)
 	\, .
\end{equation}
Up to $\mathcal{O}(\alpha)$ in vacuum, one can remove the non-trivial part of 
$q_{\scr T}$ and $c_{\scr T}^2$ by rewriting the quadratic action in terms of the following 
new variables:
\begin{equation}
\widetilde{N} 
    = 
    N \bigl( 1 + 372\alpha\kappa^4H^4 \bigr) \, , 
\qquad
\widetilde{g}_{ij}
    =
    g_{ij} 
    + 24\alpha\kappa^4 H^2 \bigl( H^2g_{ij} - 2R_{ij} \bigr)
    \, ,
\end{equation}
or in term of background and perturbation variables,
\begin{equation}
\widetilde{\overline{N}} =
	\bigl( 1+372\alpha\kappa^4 H^4 \bigr) \overline{N} \, , 
\quad \ \,
\widetilde{a} =
	\bigl( 1+12\alpha\kappa^4 H^4 \bigr) a \, , 
\quad \ \,
\widetilde{h}^{\scr TT}_{ij} = 
	\Bigl(1 - 24\alpha\kappa^4 H^2\frac{k^2}{a^2} \Bigr)
		h^{\scr TT}_{ij}
 	\, ,
\end{equation}
where 
\begin{equation}
 R_{ij} = -\frac{1}{2}\delta^{kl}\partial_k\partial_l h^{\scr TT} _{ij} 
 	+ \mathcal{O}(\epsilon^2)\,.
\end{equation}
%

%%%%%%%%%%%%%%%%%%%%%%%%%%%%%%%%%%%%%%%%%%
%%%%%%%%%%%%%%%%%%%%%%%%%%%%%%%%%%%%%%%%%%
%%%	 BIBLIOGRAPHY
%%%%%%%%%%%%%%%%%%%%%%%%%%%%%%%%%%%%%%%%%%
%%%%%%%%%%%%%%%%%%%%%%%%%%%%%%%%%%%%%%%%%%

\end{document}